\documentclass[journal]{IEEEtran}

\usepackage{amsmath,amsfonts}
\usepackage{algorithmic}
\usepackage{algorithm}
\usepackage{array}
\usepackage[caption=false,subrefformat=parens]{subfig}
\usepackage{textcomp}
\usepackage{stfloats}
\usepackage{url}
\usepackage{verbatim}
\usepackage{graphicx}
\usepackage{cite}

\usepackage{makecell,multirow,diagbox,booktabs}
\usepackage{enumerate}

\usepackage[english]{babel}
\usepackage{amssymb}
\usepackage{xspace}
\usepackage{multirow}
\usepackage{float}
\usepackage{flafter}
\usepackage{comment}

\usepackage{nicefrac}
\usepackage[acronym]{glossaries} 

\usepackage[normalem]{ulem}
\usepackage{booktabs}
\usepackage{placeins}
\usepackage{hyperref}
\usepackage[nameinlink]{cleveref} 

\usepackage{cite}

\graphicspath{{figures/}}
\DeclareGraphicsExtensions{.pdf,.jpeg,.png}

\newcommand{\platformiwl}{Intel IWL5300}
\newcommand{\platformax}{Intel AX210}
\newcommand{\platformqca}{Qualcomm Atheros QCA}
\newcommand{\platformusrp}{USRP x310}

\newcommand{\iwlshort}{IWL5300}

\newcommand{\qcashort}{QCA}

\newcommand{\iwl}{\emph{iwl5300}}
\newcommand{\intelax}{\emph{ax210}}
\newcommand{\qca}{\emph{qca}}
\newcommand{\espone}{\emph{ESP1}}
\newcommand{\esptwo}{\emph{ESP2}}
\newcommand{\esp}{\emph{ESP32}}
\newcommand{\asusone}{\emph{asus1}}
\newcommand{\asustwo}{\emph{asus2}}
\newcommand{\asus}{\emph{ASUS}}
\newcommand{\usrp}{\emph{x310}}

\newcommand{\wifi}{Wi-Fi}

\newcommand{\eg}{e.g.,~}

\providecommand{\Description}[1]{}

\newacronym{csi}{CSI}{Channel State Information}
\newacronym{cfr}{CFR}{Channel Frequency Response}
\newacronym{lti}{LTI}{linear time-invariant}
\newacronym{ofdm}{OFDM}{Orthogonal Frequency Division Multiplexing}
\newacronym{pdp}{PDP}{Power Delay Profile}
\newacronym{tof}{ToF}{time of flight}
\newacronym{rssi}{RSSI}{Received Signal Strength Indicator}
\newacronym{bfi}{BFI}{Beamforming Feedback Information}
\newacronym{cots}{COTS}{Commercial Off-The-Shelf}
\newacronym{phy}{PHY}{Physical Layer}
\newacronym{agc}{AGC}{Automatic Gain Control}
\newacronym{rf}{RF}{Radio Frequency}

\newacronym{sto}{STO}{Sampling Time Offset}
\newacronym{cfo}{CFO}{Carrier Frequency Offset}
\newacronym{csd}{CSD}{Cyclic Shift Diversity}

\newacronym{nic}{NIC}{Network Interface Card}
\newacronym{sdr}{SDR}{Software-Defined Radio}

\newacronym{std}{SD}{Standard Deviation}
\newacronym{iqr}{IQR}{Interquartile Range}
\newacronym{outratio}{OR}{Outlier Ratio}
\newacronym{mi}{MI}{Mutual Information}

\newacronym{fft}{FFT}{Fast Fourier Transform}
\newacronym{mse}{MSE}{Mean Squared Error}
\newacronym{precdev}{MPRD}{Mean Precoding Response Deviation}
\newacronym{bca}{BCa}{bias-corrected and accelerated}
\newacronym{har}{HAR}{Human Activity Recognition}
\newacronym{ota}{OTA}{over-the-air}

\newacronym{cv}{CV}{cross-validation}
\newacronym{ls}{LS}{least-squares}

\usepackage{eso-pic}
\newcommand{\IEEEacceptednotice}{%
\AddToShipoutPictureBG*{%
  \AtPageUpperLeft{%
    \raisebox{-0.55in}{%
      \makebox[\paperwidth][c]{%
        \begin{minipage}{0.93\paperwidth}
          \centering\footnotesize
          This article has been accepted for publication in \textit{IEEE Internet of Things Journal}.
          This is the author's version which has not been fully edited and content may change prior to final publication.
          Citation information: DOI
          \href{https://doi.org/10.1109/JIOT.2026.3667804}{10.1109/JIOT.2026.3667804}.
        \end{minipage}%
      }%
    }%
  }%
}%
}

\begin{document}


\title{Same Signal, Different Story: Demystifying Receiver Effects in Wi-Fi Channel State Information}

\author{Fabian Portner, 
        Francesco Gringoli, 
        Matthias Hollick, 
        and Arash Asadi%
\thanks{Fabian Portner and Arash Asadi are with Delft University of Technology, Delft, The Netherlands (email: fportner@tudelft.nl; aasadi@tudelft.nl).}%
\thanks{Francesco Gringoli is with the University of Brescia, Brescia, Italy (email: francesco.gringoli@unibs.it).}%
\thanks{Matthias Hollick is with the Technical University of Darmstadt, Darmstadt, Germany (email: matthias.hollick@seemoo.tu-darmstadt.de).}%
}

\IEEEacceptednotice

\maketitle

\pagenumbering{arabic}

\makeatletter
\def\ps@IEEEtitlepagestyle{
  \def\@oddfoot{\mycopyrightnotice}
  \def\@evenfoot{}
}
\def\mycopyrightnotice{
  {\footnotesize
  \begin{minipage}{\textwidth}
  \centering
  Copyright~\copyright~2026 IEEE. Personal use of this material is permitted. However, permission to use this \\
  material for any other purposes must be obtained from the IEEE by sending a request to pubs-permissions@ieee.org.
  \end{minipage}
  }
}

\begin{abstract}
\wifi{} sensing has emerged as a versatile tool for tasks such as localization, gesture recognition, and vital-sign monitoring, enabling applications from smart environments to personalized healthcare.
However, sensing accuracy often significantly degrades when pretrained models are deployed across different commodity receivers. We present the first systematic comparison of \gls{csi} across diverse Commercial Off-The-Shelf \wifi{} sensing platforms. Using a unified experimental setup delivering precisely precoded signals simultaneously to multiple receivers, we isolate receiver-specific variability. We find that dominant cross-device differences arise from Automatic Gain Control and consistent subcarrier nonlinearities. We propose a simple gain-alignment preprocessing step, recovering most of the lost accuracy (up to $75\%$) in cross-device Human Activity Recognition model deployments. Without preprocessing, model accuracy sharply drops---effectively breaking practical deployments. Additional analyses reveal measurable inherent differences in receiver faithfulness, sensitivity and noise. While these receiver-induced differences do not significantly affect robust sensing tasks such as Human Activity Recognition, they become relevant in scenarios demanding high precision (\eg single-shot time of flight).
Our findings demonstrate that cross-device variability in \gls{csi} is real but manageable, and we provide tools and guidelines for robust, hardware-agnostic \wifi{} sensing.
\end{abstract}

\section{Introduction}
\glsresetall[acronym]

\wifi{} \gls{csi} has evolved from a communication by-product into a sensing enabler that now powers tasks such as localization, gesture recognition, and even vital-sign monitoring. However, a critical obstacle to robust and portable sensing (that is, the ability to apply sensing models across different hardware platforms) has received limited systematic attention: \textbf{commodity receivers distort \gls{csi} in device-specific ways that degrade generalization across devices}. Differences in hardware, firmware, and signal processing pipelines, even within identical device families, cause receivers to perceive the wireless channel differently. As a result, sensing models trained on one device often fail on others, and seemingly minor discrepancies across studies may reflect hardware quirks rather than algorithmic differences. Despite the growing importance of \gls{csi}, the community lacks a clear understanding of how receiver-induced variability manifests and impacts downstream tasks.

Unlike communication systems, which rely on equalization to abstract away channel effects, \gls{csi}-based applications require a stable, interpretable relationship between the environment and the reported \gls{csi}. However, various receiver components, including the front-end, \gls{agc}, \gls{csi} estimation pipeline, and lack of synchronization, can introduce nontrivial yet often deterministic distortions, as well as differences in sensitivity and noise levels.

These effects challenge the design of portable sensing systems. In sensing, they limit model transfer. In fingerprinting, they yield device-specific signatures. In security, they introduce inconsistencies that weaken authentication. Yet no standardized methodology exists for evaluating or mitigating these differences, and many systems remain tightly coupled to specific devices.

Some effects, mainly concerning phase noise, have long been investigated and can either be corrected for or circumvented~\cite{ratnam2024optimal,niu2021wimonitor,qian2014pads,zubow2021phase,zhang2019calibrating,li2017indotrack,widrone,chen2019residual,vogt2019precise,tadayon2019decimeter,yi2024enabling,li2021kalman}.
Some works have gone further and investigated static, nonlinear receiver-induced distortions in \gls{csi}, such as subcarrier-dependent amplitude and phase artifacts~\cite{xie2015precise,zhu2018pi, jiang2021eliminating}, and shown that these effects are repeatable and thus calibratable. However, these studies examine only a narrow set of devices and provide limited insight into how such distortions manifest across platforms or affect downstream performance more broadly.

We address this gap by jointly studying receiver effects in \gls{csi} and deep learning-based sensing models through a systematic black-box comparison across \gls{cots} platforms. Using a unified experimental setup that delivers identical, precisely precoded signals to multiple receivers, we isolate and quantify receiver-induced differences. We enhance existing \gls{csi} extraction tools \cite{xie2015precise, halperincsitool, hernesp32} to support heterogeneous \glspl{nic}, enabling direct comparison of correctable distortions and inherent hardware differences.

In order to assess practical consequences, we evaluate three representative tasks: Doppler-based velocity estimation, \gls{tof} extraction from \gls{pdp}, and deep learning-based \gls{har}. We find that most cross-device differences stem from predictable effects, namely \gls{agc} behavior and subcarrier nonlinearities, which can be corrected using a lightweight gain-alignment step. After correction, residual differences in sensitivity and noise remain, but have limited impact on robust sensing tasks, becoming relevant only in precision-critical settings such as single-shot \gls{tof}. Our findings offer practical guidance and tools for building portable, hardware-agnostic \wifi{} sensing systems.

\newpage
\subsection{Contributions}

Our contributions are as follows:

\begin{itemize}
    \item We design a unified experimental setup that delivers identical precoded signals to multiple receivers simultaneously, enabling controlled black-box comparison without ground truth, and release all collection and processing code as open source.

    \item We systematically evaluate \gls{csi} quality across all major \gls{cots} \wifi{} platforms, including \textit{\platformiwl{}}, \textit{\platformax{}}, \textit{\platformqca{}}, \textit{Broadcom}, \textit{Espressif ESP32}, and a \textit{\platformusrp{}}, and publicly release the collected \gls{csi} and derived data.
    
    \item We identify that most cross-device differences stem from predictable receiver-induced distortions, including consistent nonlinear subcarrier shapes and \gls{agc} behavior, which can be effectively corrected.

    \item We propose a simple gain removal with feature-wise standardization that outperforms alternative schemes and recovers up to $75\%$ of the cross-device accuracy loss for several \gls{har} models.

    \item We further quantify residual differences in noise, faithfulness, and sensitivity, showing that while measurable, these have limited impact on practical tasks such as \gls{har} and Doppler-based velocity estimation.
\end{itemize}

\section{Background} \label{sec:background}
\subsection{Channel State Information} \label{subsec:csi-bg}

The wireless channel between a transmitter and receiver varies as people move, objects reflect, and paths change. To ensure reliable communication, receivers repeatedly estimate the channel state and compensate for its variations.

\wifi{} employs \gls{ofdm}, dividing the communication bandwidth into multiple narrowband subcarriers. Within the duration of a single packet, the channel is typically assumed linear and time-invariant. Under this assumption, the received \gls{ofdm} symbol on subcarrier~$k$ can be expressed simply as
\begin{align} \label{eq:freq-symbol-model}
R_k = H_k S_k,
\end{align}
where $S_k$ denotes the transmitted symbol, $R_k$ the received symbol, and $H_k$ the \gls{cfr}. In multipath environments, $H_k$ decomposes into a sum of $P$ propagation paths~\cite{tse2005fundamentals}:
\begin{align*}
H_k = \sum_{p=1}^P \alpha_{p,k} \exp\left(-j 2\pi f_k \tau_p \right),
\end{align*}
where $\alpha_{p,k}$ and $\tau_p$ represent the attenuation and delay of the $p$-th path, and $f_k$ is the frequency of subcarrier~$k$. Over multiple packet transmissions, these path parameters evolve slowly, causing $H_k$ to vary.

Receivers estimate the channel from fixed training symbols in the physical-layer preamble, as defined in the 802.11 standard, yielding per-packet \gls{csi} vectors $\hat{H}[n] = [\hat{H}_1[n], \dots, \hat{H}_K[n]]$. Ideally, these estimates match the true channel up to a constant scaling factor ($\hat{H}[n] = c H[n]$). In practice, \gls{cots} receivers introduce distortions due to gain control, synchronization offsets, and hardware nonlinearities, which we model as a receiver-specific transformation $\hat{H}[n] = E(H[n])$. For each subcarrier, we decompose $E$ into amplitude and phase distortions:
\begin{align*}
\hat{H}_k[n] = \beta_k[n] \exp\left(-j 2\pi \phi_k[n] \right) H_k[n] + \eta_k[n],
\end{align*}
where $\beta_k[n]$ represents amplitude effects, $\phi_k[n]$ phase distortions, and $\eta_k[n]$ additive noise. For tractability, $\beta_k[n]$ and $\phi_k[n]$ are typically modeled explicitly. Phase distortions are often taken to be linear in subcarrier index~$k$ due to \gls{sto}, \gls{cfo}, and \gls{csd}~\cite{wu2024sensing, zhu2018pi}, while gain control is approximated as a global amplitude factor across subcarriers.

Real-world receivers, however, exhibit additional artifacts, including nonlinear subcarrier responses and chipset-specific quirks, that violate these assumptions. Prior work~\cite{zhu2018pi, jiang2021eliminating} shows that such nonlinearities are static and repeatable, enabling calibration. To explicitly include them, we model
\begin{align} \label{eq:amp-phase-model}
\beta_k[n] &= G[n] \cdot A_k, \nonumber \\
\phi_k[n] &= a[n] \cdot k + b[n] + \Phi_k,
\end{align}
where amplitudes factor into a global gain $G[n]$ and subcarrier profile $A_k$, and phases into a linear trend plus static offset $\Phi_k$. This formulation captures dominant receiver-side distortions.

\subsection{Related Work} \label{subsec:related}

\noindent\paragraph{CSI Extraction on Commodity Devices}
A growing body of tools enables \gls{csi} collection on commodity \wifi{} hardware. The pioneering Intel 5300 \gls{csi} Tool~\cite{halperincsitool} introduced modified firmware and drivers to obtain \gls{csi} from 802.11n \glspl{nic}. Subsequent platforms expanded \gls{csi} access: the Atheros \gls{csi} tool~\cite{xie2015precise} targeted Qualcomm Atheros \glspl{nic}, the Nexmon framework~\cite{nexmon} provided extraction for Broadcom 802.11ac chips, Espressif’s \esp{} \wifi{} module~\cite{hernesp32} made low-cost boards popular, and the PicoScenes platform~\cite{jiang2021eliminating} unified support for Intel, Atheros, and \gls{sdr} devices with additional capabilities. These tools have opened \gls{csi} research to a variety of devices, but each receiver’s \gls{csi} may embody unique quirks. Our work complements this landscape with a side-by-side, cross-device analysis of \gls{csi} from multiple \gls{cots} receivers using a common transmitter to pinpoint receiver-induced distortions.

\noindent\paragraph{Receiver-Induced Distortions} Previous studies have highlighted that \gls{csi} measurements suffer from hardware-induced nonlinearities varying across chipsets. Zhuo et al.\,\cite{zhuo2016ident} observed nonlinear distortions they attributed to IQ imbalance, later introducing $\pi$-Splicer \cite{zhu2018pi} to correct them. Jiang et al.\,\cite{jiang2021eliminating} re-examined those distortions, attributing them to baseband filter artifacts, and comparing across Intel \iwlshort{} and Atheros \glspl{nic}. Chi et al.\,\cite{widrone} applied a simple cable-based calibration to eliminate these distortions, assuming temporal consistency.

However, these prior studies each examined a limited set of devices and the origin of nonlinear distortions, whereas our work provides a broader, side-by-side comparison across diverse receivers and evaluates the practical impact of such distortions on sensing applications.

\section{Setup and Methodology} \label{sec:setup}

To characterize receiver-dependent effects, we must isolate them from the variability of wireless channels. In \gls{ota} scenarios, fluctuating channel conditions and transmitter-induced effects make \gls{csi} difficult to interpret because no channel ground truth is available. We therefore build a controlled testbed and introduce a precoding methodology that emulates known, time-varying channel changes.

We leverage a high-fidelity \gls{sdr} as our transmitter, which minimizes TX-induced variability and enables precise control over the transmitted \wifi{} frames. By routing the signal over cable and distributing an identical input to all receivers via a low-noise power splitter, we eliminate wireless channel variability and can directly compare how different receivers respond to deliberate, controlled channel modifications.

\subsection{Experimental Setup}

\begin{figure}[!t]
    \centering
    \includegraphics[width=0.9\linewidth]{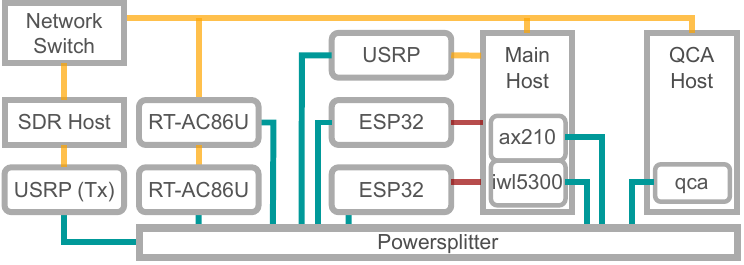}
    \par\vspace{1mm}
    \begin{minipage}[b]{0.45\linewidth}
      \centering
        \includegraphics[width=\linewidth]{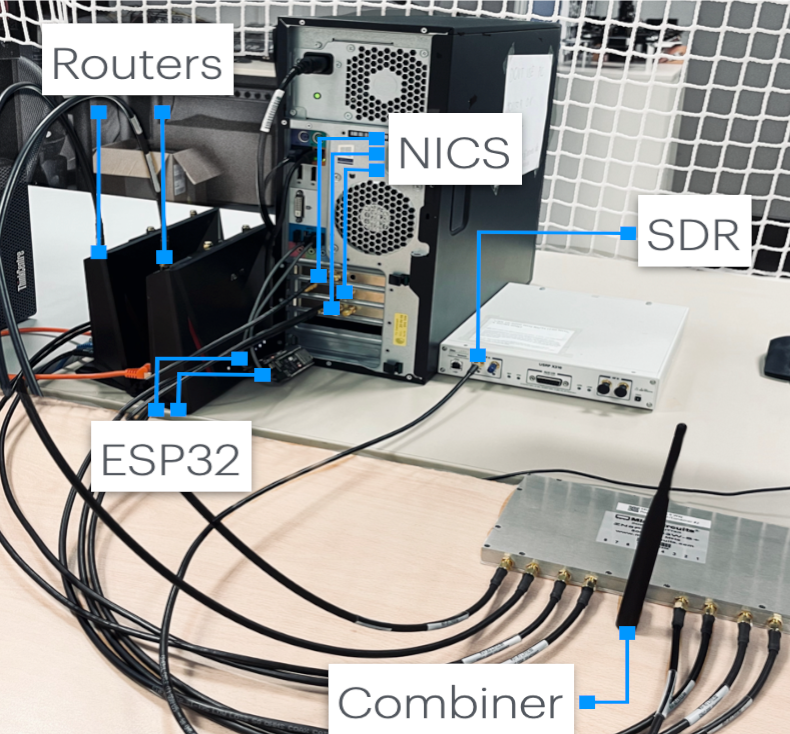}
    \end{minipage}
    \begin{minipage}[b]{0.45\linewidth}
      \centering
        \raisebox{-3.5mm}{
      \includegraphics[width=\linewidth]{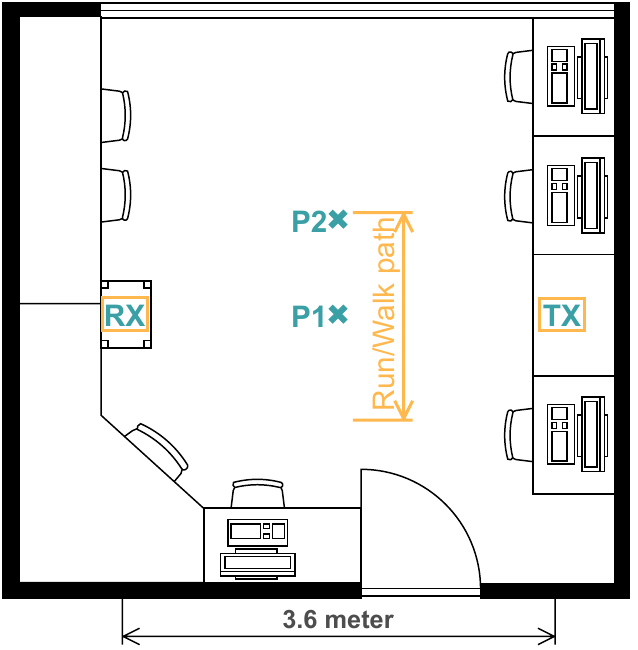}
      }
    \end{minipage}

    \caption{Experimental setup. Top: schematic representation with RF connections in cyan, ethernet in yellow, USB in red. Bottom: photograph of the physical setup and corresponding floorplan with \gls{har} person positions.}
    \Description{Top: schematic showing eight receivers connected via power splitter to a single transmitter. Bottom: photo of the real setup and floorplan showing the same hardware arrangement.}
    \label{fig:experiment}
\end{figure}

To compare \gls{csi} across multiple receivers, the experimental setup includes eight devices spanning six distinct receiver models with descriptive names used in the rest of the paper:

\begin{itemize}
    \item \textbf{2\texttimes ASUS RT-AC86U} routers named \emph{\asusone{}} and \asustwo{}
    \item \textbf{2\texttimes ESP32-S3} microcontrollers named \emph{\espone{}} and \esptwo{}
    \item \textbf{1\texttimes Intel AX210} \gls{nic} named \intelax{}
    \item \textbf{1\texttimes Intel IWL5300} \gls{nic} named \iwl{}
    \item \textbf{1\texttimes USRP N2954-R} \gls{sdr} named \usrp{}
    \item \textbf{1\texttimes Qualcomm Atheros QCA AR9462} named \qca{}
\end{itemize}

To assess intra-family variability, we duplicate the least and most constrained \gls{cots} devices: the high-end \asus{} RT-AC86U and the low-cost \esp{}. We observe minor but consistent differences between such devices (see \Cref{sec:cross-sensing}), indicating that receiver-induced effects can arise even within a single hardware family. To limit setup complexity, we restrict duplication to these two representative cases and leave a more exhaustive intra-family analysis for future work.

The Intel cards and \usrp{} are operated by an \emph{Experiment Host PC} running Ubuntu~$22.04$, and configured via PicoScenes~\cite{jiang2021eliminating}. The ESPs are attached to the same host and use a custom-developed extractor for scripted \gls{csi} collection, while \gls{csi} from the \asus{} routers is obtained via a Nexmon-based custom extractor~\cite{nexmon}. A second PC running Ubuntu~24.04 operates the \qca{} independently to avoid driver conflicts with PicoScenes\footnote{PicoScenes officially features Atheros devices, but this is currently disabled due to \href{https://gitlab.com/wifisensing/picoscenes-issue-tracker/-/issues/197}{Issue 197}.}, and we extract \gls{csi} using a ported version of the Atheros \gls{csi} tool~\cite{xie2015precise} compatible with newer kernels.

For the transmitter, another PC operates a \emph{second} USRP N2954-R. We use MATLAB, specifically the WLAN Toolbox, to generate IQ samples corresponding to $802.11n$ \wifi{} frames. 
The transmitter connects via cable to a waveguide power splitter (\href{https://www.minicircuits.com/WebStore/dashboard.html?model=ZN8PD1-63W-S%2B}{ZN8PD1-63W-S+}), which distributes the same signal to all eight receivers. A schematic overview of our setup is presented in \Cref{fig:experiment}.

The iwl5300 and Atheros \gls{csi} tools are incompatible for collecting \gls{csi} from the same frames: the former reports \gls{csi} only for frames addressed to a hardcoded MAC address, while the latter requires frames addressed to its hardware MAC address. We therefore send interleaved frames that differ only in the destination MAC address. The other extractors have no such restriction; for them we simply filter frames addressed to the \iwl{}.

We conduct all experiments using single-antenna configurations and a channel bandwidth of $20$~MHz, as imposed by the ESPs, which operate only on $20$~MHz in the $2.4$~GHz band and have a single antenna. All devices operate passively in monitor mode, without association to an access point.

To ensure uniform analysis, we transform \gls{csi} from all devices into a common format and store it in Apache Parquet files. The Experiment Host coordinates all devices over a local network for synchronized operation and centralized data collection, and the pipeline is scripted to detect and recover from common runtime errors.

\subsection{OFDM Precoding} \label{subsec:precoding}

In \gls{ota} experiments, assessing \gls{csi} accuracy is difficult because no channel ground truth is available. We therefore exploit the fact that the transmitted OFDM symbols $x_k$ are fixed by the 802.11 standard and introduce controlled modifications via precoding. Using the MATLAB WLAN Toolbox, we generate \wifi{} $802.11n$ frames and precode all symbols in a packet with subcarrier-specific complex scaling factors $c_k \in \mathbb{C}$, yielding $x'_k = c_k x_k$. From the receiver’s perspective, this is indistinguishable from a change in the channel, and the received signal becomes $y = H \odot c \odot x$, where $H$ is the true channel and $\odot$ denotes the Hadamard product.

Because the receiver does not know $c$, it effectively estimates $\hat{H} = E(H \odot c)$. If the channel remains constant, $c$ appears as a consistent, relative change in $\hat{H}$, providing a stable reference to assess how closely the estimates follow the expected changes and how much distortion the internal estimation pipeline $E$ introduces. By precisely controlling $c_k$ while keeping all other conditions fixed, we isolate the behavior of $E$ and systematically evaluate how each receiver responds to small, deliberate input changes: its sensitivity, added noise, and temporal consistency.

\section{Cross-receiver sensing} \label{sec:cross-sensing}

We study how receiver hardware affects performance and portability in \wifi{} sensing by measuring gaps in \gls{har} and Doppler tasks and showing that simple preprocessing largely restores generalization across heterogeneous receivers.

\subsection{Impact of different receivers} \label{subsec:aril-ondevice}
\begin{figure}[!b]
  \centering
  \includegraphics[width=0.94\linewidth]{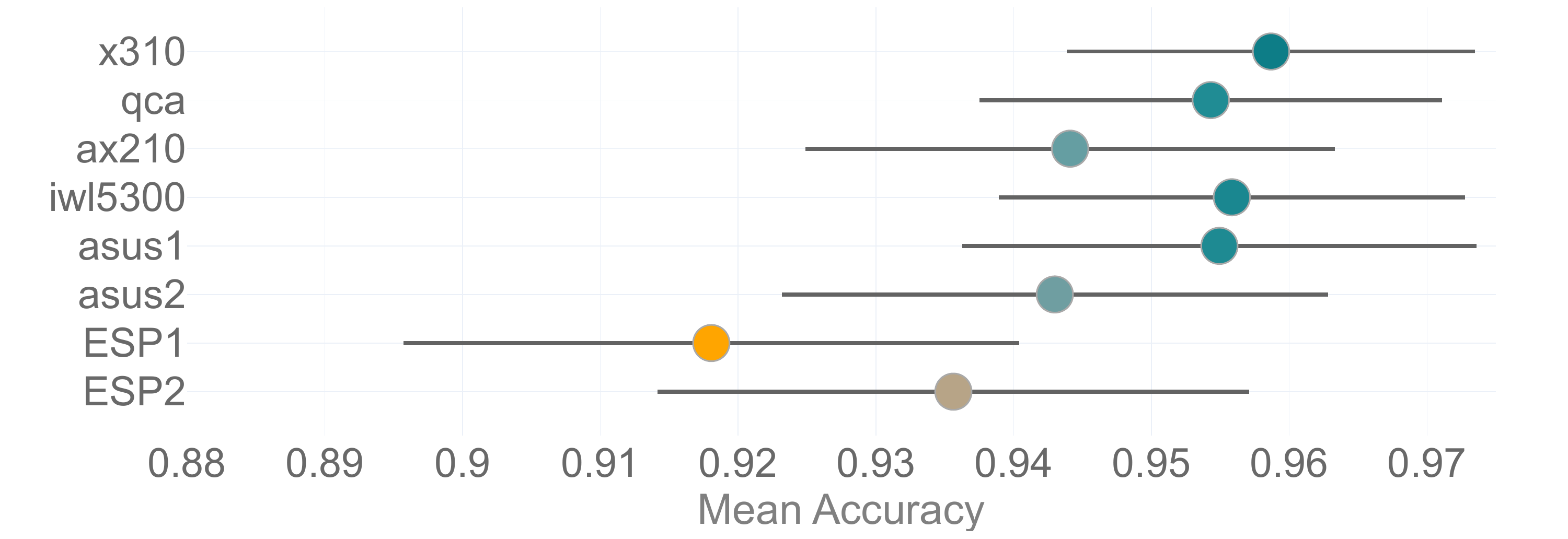}
  \caption{Per-receiver accuracy $\pm 1$ SD, estimated with $20 \times 5$-fold \gls{cv}.}
  \Description{Mean accuracy with standard-deviation whiskers for each receiver, obtained from repeated 5-fold \gls{cv}.}
  \label{fig:aril-unscaled-ondevice} 
\end{figure}

We first ask whether hardware differences meaningfully affect sensing tasks such as \gls{har}. To address this, we use ARIL~\cite{aril}, an open-source \gls{har} model with a ResNet backbone, and test it on data from different receivers to evaluate device-dependent sensitivities in a neural-network application. We follow the authors' published training procedure\footnotemark[1], but backpropagate only an activity-classification loss (ignoring the location head). Because \gls{har} requires human motion in an actual propagation environment, we replace the transmitter--power splitter cable with a TX/RX antenna pair.

\paragraph*{Dataset and protocol}

A single participant provided informed consent and performed seven activities facing the RX (jumping, standing still, walking, running, squat, clapping, and waving arms) plus an empty-room baseline. The participant stood in either LOS or NLOS (\Cref{fig:experiment}), yielding $90$ repetitions per activity across conditions. We record at $500$ fps for $3$ s and subsample each series to $700$ frames of $56$ subcarriers to mitigate packet-loss inhomogeneity. We focus on receiver variability: in one office setup, we vary LOS/NLOS positions and repeat trials, while all receivers see the same channel realizations via the power splitter. This suffices for our purpose, since we use \gls{har} to probe task-relevant information exposure per device, not to benchmark \gls{har} in general. As in the original implementation, we do not apply data standardization.

We evaluate model performance using \textbf{stratified $5$-fold \gls{cv} repeated $20$ times} ($100$ train/test splits per receiver). Repeated $k$-fold \gls{cv} is recommended for small–moderate datasets as a low-bias, lower-variance estimator~\cite{kfoldrecommendation,nounbiasedkfold,kim2009estimating}. We report mean accuracy and its standard deviation across splits.

\paragraph*{Results}
\Cref{fig:aril-unscaled-ondevice} reveals that (\qca{}, \iwl{}, \asus{}) consistently achieve accuracies around $95\%$, whereas the two \esp{} modules lag by about $2$--$4\%$. Interestingly, \texttt{\espone{}} underperforms \texttt{\esptwo{}}, despite nominally identical specification, suggesting hardware variability within the \esp{} family that could merit further investigation in future work.

\paragraph*{Key takeaway}
While all devices enable ARIL to achieve high accuracy, hardware differences between receivers meaningfully affect classification performance. Employing a low-bias estimator such as repeated stratified \gls{cv} is essential for quantifying these effects and comparing improvements in \wifi{} sensing pipelines.

\begin{figure}[!t]
  \centering
  \includegraphics[width=0.9\linewidth]{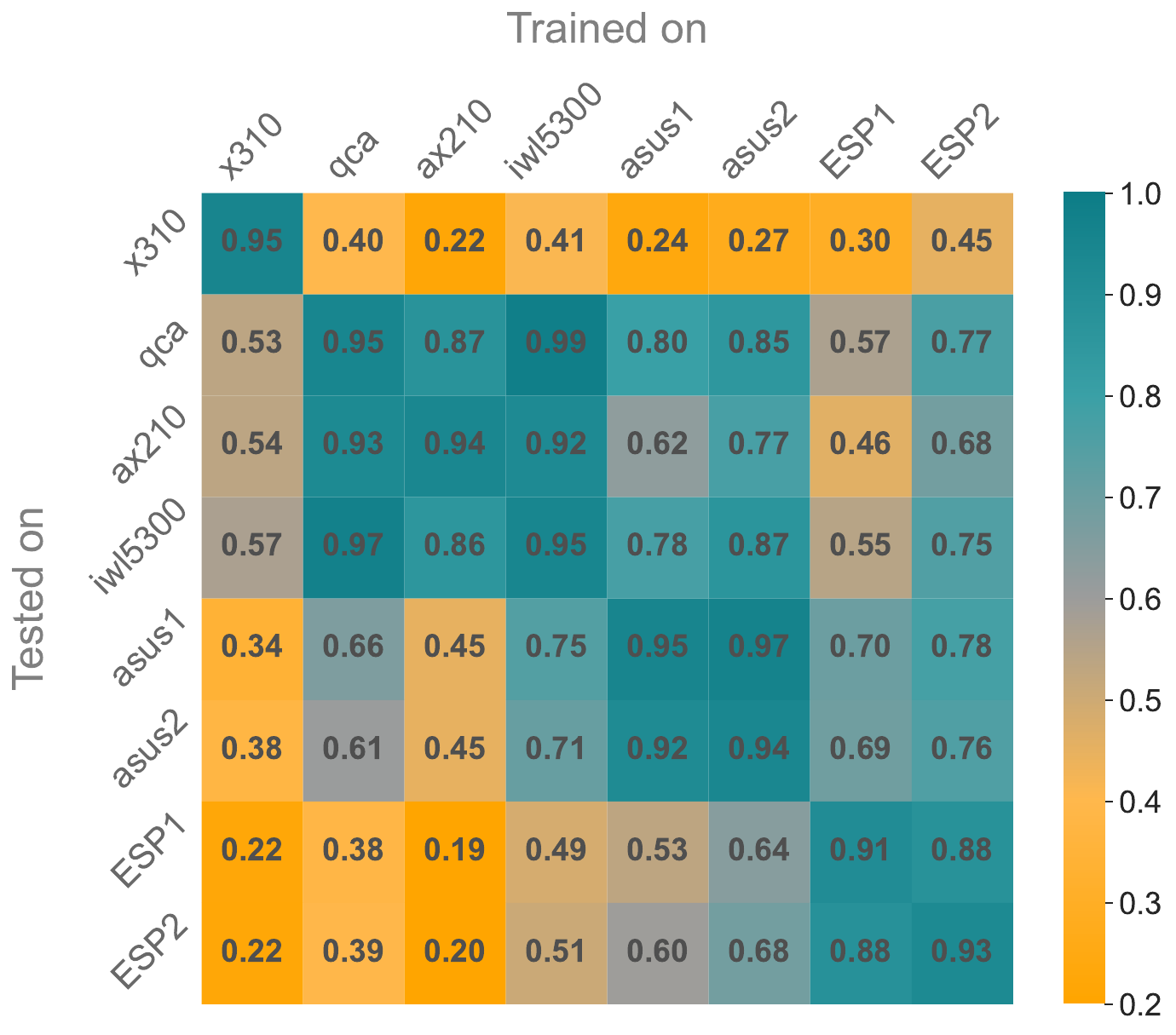}
  \caption{Confusion matrices when ARIL is trained on one receiver (rows) and tested on others (columns), without normalization.}
  \Description{Confusion matrices when ARIL is trained on one receiver (rows) and tested on others (columns), without normalization.}
  \label{fig:aril-unscaled-confusion}
\end{figure}

\subsection{When the receiver changes: cross-device generalization} \label{subsec:cross-device}

\footnotetext[1]{\href{https://github.com/geekfeiw/ARIL}{ARIL GitHub repository.}}

Retraining models for every receiver does not scale: labeling new data is costly, training adds overhead, and both steps hinder deployment at scale. As \wifi{} sensing moves toward broader use, reusing pretrained models across devices becomes a \textit{practical necessity}. Yet while cross-environment transfer has been studied, the role of the receiver itself has gone largely unexamined.

\paragraph*{Cross-device evaluation}
We now ask whether a model trained on one receiver transfers to others. Because different devices report \gls{csi} on very different amplitude scales, we add a single feature-wise z-score standardization step, separately for each receiver and for the train and test sets; all other hyperparameters and training choices remain identical to the original ARIL implementation. When training and testing on the same receiver, we keep an $80/20$ train--test split. For cross-device transfer, we instead train ARIL on all data from one device and test on all samples from another, so these evaluations benefit from a slightly larger training set; \Cref{fig:aril-unscaled-confusion} shows the resulting confusion matrices. Performance drops significantly when classifying on data from a different receiver. The most severe degradation occurs when training on \usrp{}: although it performs well on its own data, accuracy drops below $60\%$ on every other device, reaching as low as $20\%$ on \texttt{\espone{}}. \texttt{\qca{}}, \texttt{\intelax{}} and \texttt{\iwl{}} exhibit good transfer accuracies, likely aided by the increased training data. The two ESPs generalize well across each other, with only a slight $3\% - 5\%$ accuracy reduction, but model performance breaks across device families.

\begin{figure*}[!ht]
    \centering
    \includegraphics[width=0.9\textwidth,alt={AGC scaling.}]{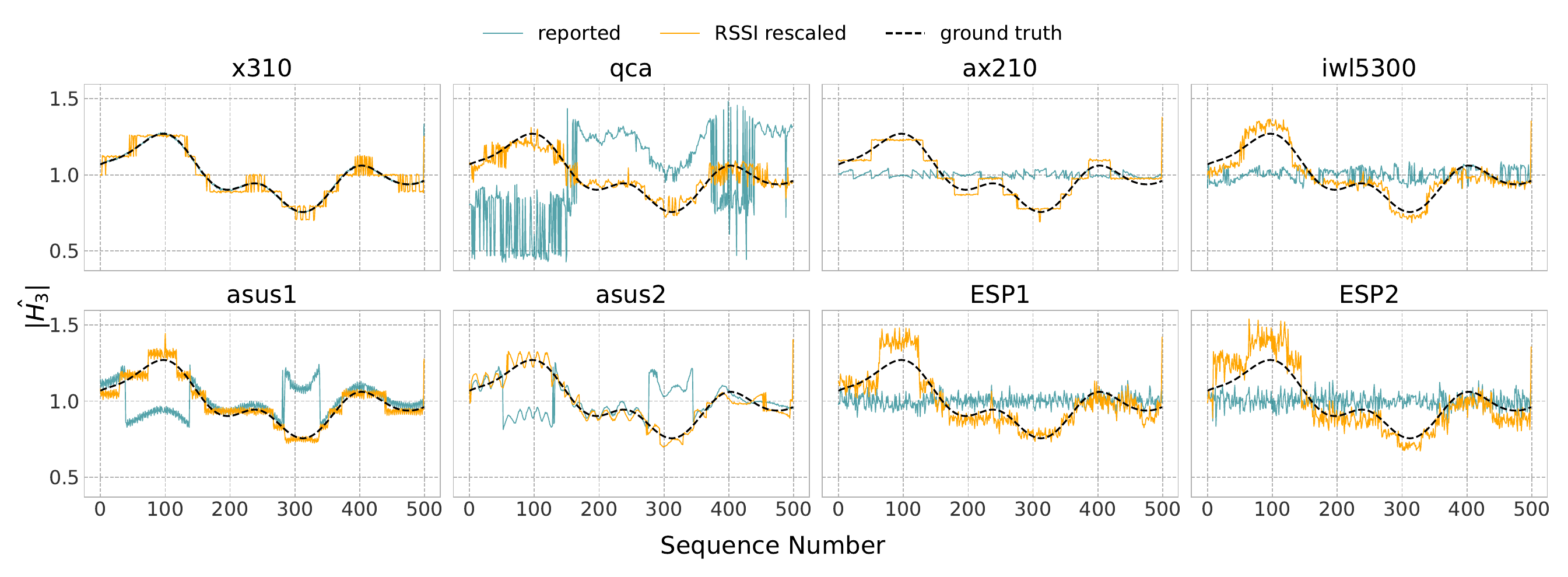}
    \caption{Reported \gls{csi} amplitudes of subcarrier $k=3$ per receiver, normalized to series mean, versus ground truth scaling curve.}
    \Description{Reported \gls{csi} amplitudes of subcarrier $k=3$ per receiver, normalized to series mean, versus ground truth scaling curve, showcasing vastly different \gls{AGC} behavior across devices.}
    \label{fig:gain-control}
\end{figure*}

\subsection{A likely cause: AGC} \label{subsec:agc-over-cable}

\gls{har} works by recognizing patterns of relative \gls{csi} change over the activity duration. A well-known confounder over time is \gls{agc}, which dynamically rescales \gls{csi} amplitudes based on received signal strength. This introduces an unknown scaling effect that obscures amplitude variations. Without a gain indicator, it is challenging to distinguish channel-induced changes from \gls{agc} artifacts, hindering applications that rely on accurate amplitude information. Because amplitude directly influences downstream tasks (\eg Doppler extraction), \gls{csi}-based applications should preserve its natural variation.

One practical mitigation is to rescale \gls{csi} amplitudes using \gls{rssi} via the factor\cite{gao2020crisloc}
\begin{align}\label{eq:rssi-rescale}
    s = \sqrt{\frac{10^{\nicefrac{RSS}{10}}}{\sum_k |\hat{H}_k|^2 }}.
\end{align}
However, \gls{rssi} is a coarse metric with much lower resolution than \gls{csi}, limiting its effectiveness.

To investigate gain control across devices, we again connect the transmitter and power splitter by cable and apply a common amplitude precoding factor across all subcarriers (see \Cref{subsec:precoding}), varying transmit gain smoothly over time. In an ideal scenario without \gls{agc}, a single global normalization would let each subcarrier amplitude follow the expected curve.

\Cref{fig:gain-control} shows \gls{csi} amplitude for subcarrier $k=3$ before and after \Cref{eq:rssi-rescale}. The results reveal stark differences in how gain control affects various devices. The \usrp{} (no \gls{agc}) accurately preserves the original variations; the measured curve shadows the ground truth. In contrast, \asus{} devices exhibit large, infrequent step changes, making gain control effects easy to spot. Intel devices also show step-like adjustments, albeit smaller and more frequent, making them harder to discern visually. For \esp{} and \qcashort{} devices, the impact of gain control is indistinguishable from noise; its application appears almost random, completely obscuring the underlying pattern to the human eye.

While \gls{rssi} rescaling moves amplitudes toward the expected curves, it still leaves substantial device-to-device variation because \gls{rssi} is itself receiver-dependent.

\subsection{No gain, no pain}
To address this, we propose to simply remove large-scale gain by a simple, yet effective normalization
\begin{align} \label{eq:amp-norm}
\hat{H}^N_k &= \frac{\hat{H}_k}{\frac{1}{K} \sum_{k} |\hat{H}_k|}.
\end{align}
The impact of this transformation on the information content can be stated precisely by investigating the Fisher information~\cite{lehmann2006theory}.
Under a differentiable transformation $\mathbf{y} = g(\mathbf{h})$, the Fisher information matrix becomes $\mathcal{I}_g = J\, \mathcal{I}\, J^\top$,
where $J = I - \frac{1}{K} \mathbf{1} \mathbf{1}^\top$ is the Jacobian of the normalization. This projection removes the component along the constant-gain direction. The rank of the Fisher information consequently drops by (just) one $\text{rank}(\mathcal{I}_g) = \text{rank}(\mathcal{I}) - 1$.
Contrary to some prior claims~\cite{ratnam2024optimal}, this discards not all amplitude information but a single scalar: the large-scale gain \emph{shared across all subcarriers}. However, motion and environmental changes can affect large-scale gain, so this transformation may still suppress signal the network could otherwise exploit. The question is \textit{whether the remaining dimensions, those orthogonal to global gain, still contain enough information for accurate sensing}.

\paragraph*{Retraining and outcome}

\begin{figure}[!b]
  \centering
  \includegraphics[width=0.9\linewidth]{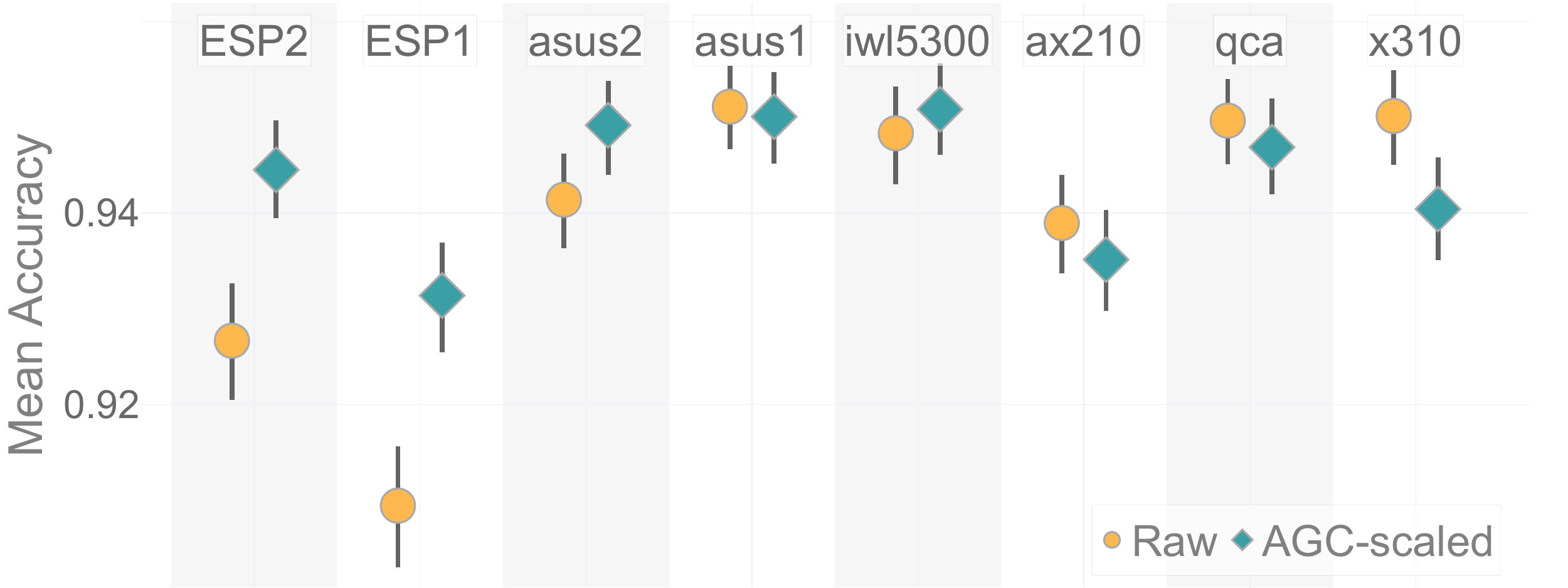}
  \caption{Per-receiver mean accuracy before/after $\ell_1$ normalization. Error bars: $99\%$ \gls{bca} bootstrap CIs from $10\,000$ resamples of the $K=100$ \gls{cv} folds.}
  \Description{Mean classification accuracy for each receiver before and after $\ell_1$ normalization. Error bars show $99\%$ \gls{bca} bootstrap confidence intervals, computed from $10\,000$ resamples of the \gls{cv} folds for each receiver.}
  \label{fig:aril-agc-comparison}
\end{figure}

To test this, we retrain ARIL on the normalized data under the setup presented in \Cref{subsec:cross-device}. For each receiver, we quantify uncertainty in the mean accuracy via nonparametric bootstrapping: we draw $10\,000$ resamples (with replacement) of the $K=100$ \gls{cv} folds, compute the mean accuracy for each resample, and report the $0.5$--$99.5$ \gls{bca} percentile limits~\cite{efron1987}, which adjust for median bias and skew. These intervals indicate whether the observed changes in mean accuracy are compatible with random variation. As shown in \Cref{fig:aril-agc-comparison}, performance is mostly maintained or improved. The \esp{} devices and \asusone{} benefit most, likely due to stabilization of erratic gain behavior, whereas the \usrp{} exhibits a small drop, indicating that without \gls{agc}, absolute gain carried some useful signal.

\begin{figure}[!t]
  \centering
  \includegraphics[width=0.9\linewidth]{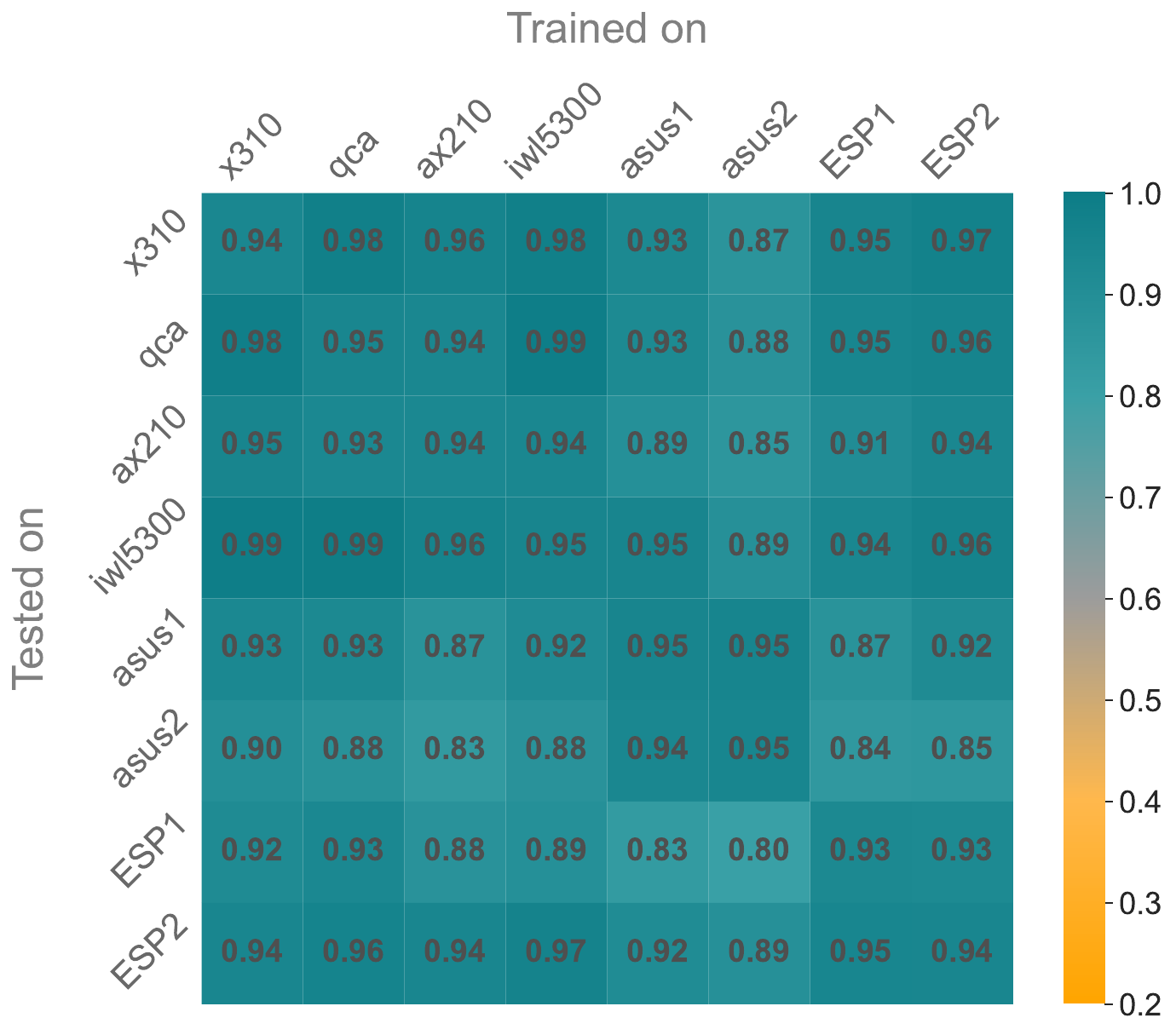}
  \caption{Confusion matrices after $\ell_{1}$ normalization, showing markedly improved cross-device generalization.}
  \Description{Confusion matrices after $\ell_{1}$ normalization, showing markedly improved cross-device generalization.}
  \label{fig:aril-scaled-confusion}
\end{figure}

\paragraph*{Cross-device generalization improves}

\Cref{fig:aril-scaled-confusion} shows updated confusion matrices. Normalization yields a substantial boost in cross-device performance: all off-diagonal accuracies now exceed $82\%$, most exceeding $95\%$ and some nearing $99\%$. A simple rescaling step effectively restores generalization in a way that preserves performance across all devices. We note that nonlinearity equalization did not yield any notable improvement, indicating that learned features rely more on relative changes over time. This shows that the shift in distribution between different receivers is mainly due to \gls{agc}.

\paragraph*{Outlook}
Normalization resolves much of the observed divergence, but not all. Worst-case, accuracy still drops by up to $10$ - $15\%$. For industrial deployments with controlled hardware variation, this may be acceptable. But in scenarios involving heterogeneous receivers or crowdsourced datasets, additional techniques, such as multi-device training or fine-tuning, may be necessary to ensure full portability.

\subsection{Alternative gain preprocessing methods}

\begin{figure}[!b]
  \centering
  \includegraphics[width=\linewidth]{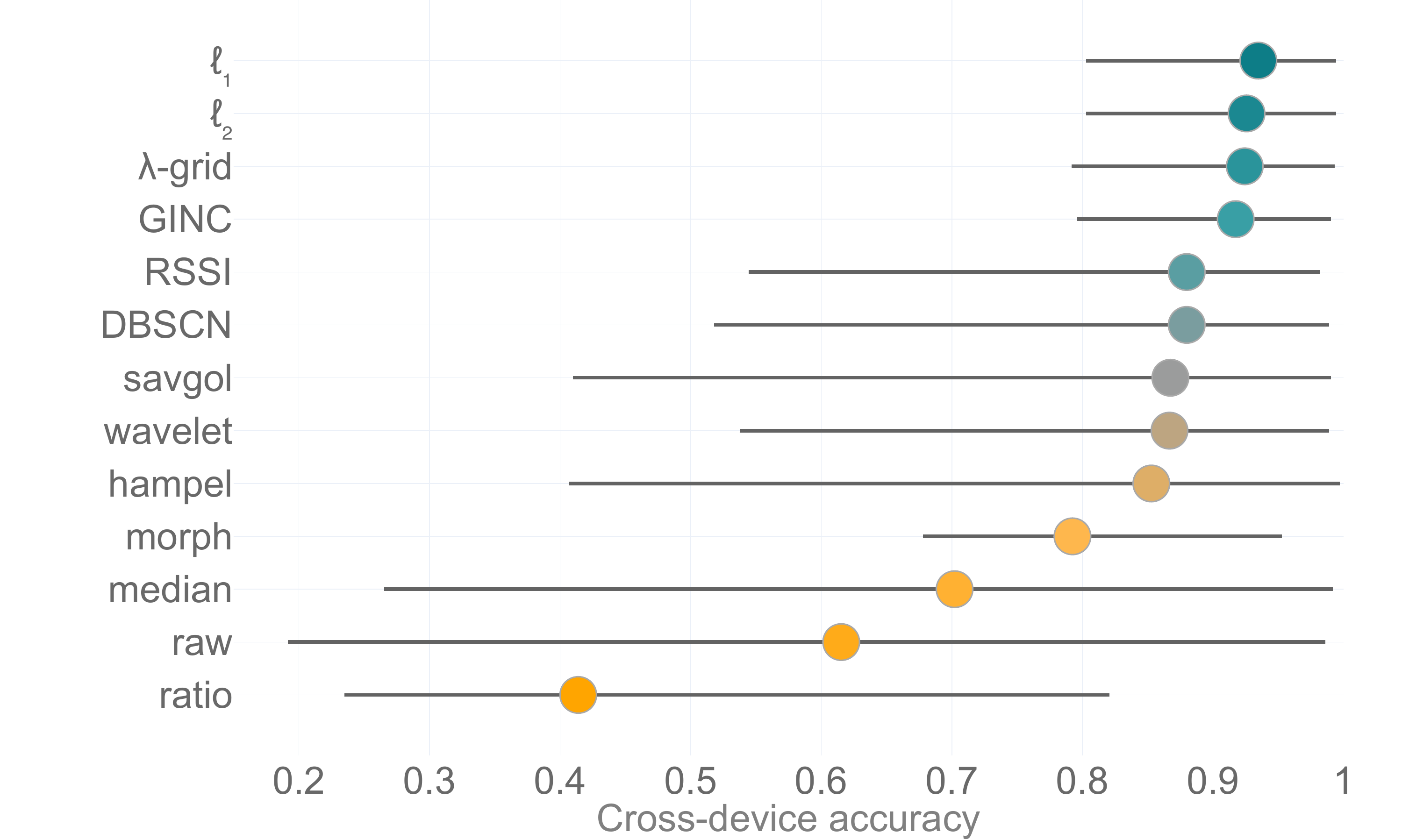}
  \caption{Cross-device performance under different preprocessing strategies; Markers show median, whiskers worst-case accuracies across all devices.}
  \Description{Cross-device performance under different preprocessing strategies; Markers signify median, whiskers worst-case accuracies across all devices.}
  \label{fig:preprocessing-comparison}
\end{figure}
We now evaluate $\ell_1$ normalization (\Cref{eq:amp-norm}) against alternative preprocessing strategies proposed to mitigate \gls{agc}-induced distortions in \gls{csi}. 

We cover three categories: (i) methods that discard large-scale gain, including $\ell_2$ normalization~\cite{gaussian2020}, gain-invariant features (morphological~\cite{chen2023} and double-ratios~\cite{yi2024enabling}); (ii) explicit gain-recovery methods, which attempt to recover a consistent amplitude scale despite \gls{agc}, including \gls{rssi}-based amplitude rescaling (\Cref{eq:rssi-rescale}), DBSCAN clustering~\cite{liu2021wiphone}, and two maximum-likelihood algorithms by Ratnam et al.~\cite{ratnam2024optimal}---GINC (Alg.\,1), and $\lambda$-grid (Alg.\,2); and (iii) smoothing filters to suppress \gls{agc} fluctuations: median\cite{wisign2017}, Savitzky–Golay~\cite{zeng2018fullbreathe}, wavelet~\cite{wisleep2014}, and Hampel~\cite{wisleep2014}.

\paragraph*{Results}
\Cref{fig:preprocessing-comparison} shows $\ell_1$ normalization outperforming all alternatives, with median cross-device accuracy above $90\%$. Explicit gain-recovery methods are competitive, trailing slightly in both median and worst-case accuracy. Smoothing-based methods drop to $85$-$89\%$ median and struggle in worst-case stability. Morphological features, median filtering, and raw \gls{csi} each stay below $80\%$ median. Double-ratio features perform worst overall, at $\approx40\%$, below raw data.

\paragraph*{Take-away}
Explicit gain-recovery offers no meaningful accuracy advantage despite higher computational cost. A simple, carefully chosen normalization---our proposed $\ell_1$ procedure---consistently yields the highest cross-device accuracy, superior robustness, and far lower compute.

\subsection{Standardization ablation}

\begin{figure}[!b]
  \centering
  \includegraphics[width=\linewidth]{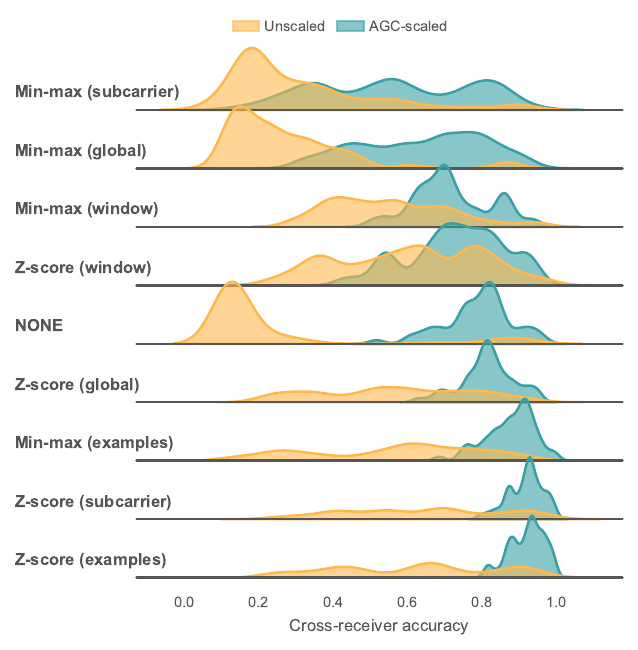}
  \caption{Distributions of cross-device accuracies under different standardization strategies, with and without gain normalization. Curves aggregate off-diagonal accuracies across all train--test device pairs.}
  \Description{Density plots of cross-device accuracies for several standardization strategies, comparing raw and gain-normalized inputs.}
  \label{fig:standardization-ablation}
\end{figure}

We now ablate classic standardization schemes that might compensate for \gls{agc}-induced scale differences. We combine min--max and z-score standardization with four scopes: (i) \emph{global}, a single transform over all CSI values; (ii) \emph{window}, one transform per example; (iii) \emph{subcarrier}, one transform per subcarrier, pooling all times and examples; and (iv) \emph{examples}, one transform per time--frequency bin across examples (feature-wise standardization).

\paragraph*{Results}
\Cref{fig:standardization-ablation} shows the resulting distributions of cross-device accuracies. Standardization clearly matters: example-wise standardization across examples (scope (iv)) gives the strongest cross-device performance, indicating that within-example amplitude changes carry useful information. Global, window, and subcarrier scopes provide smaller or inconsistent gains. Across all variants, gain-normalized inputs give better cross-device performance: accuracies are higher on average, and severe failures on particular train--test receiver pairs become much less common.

\paragraph*{Take-away}
Standardization improves cross-device robustness, but it does not replace gain normalization. The best results arise from combining example-wise standardization with our gain normalization, which together deliver the highest and most stable accuracies across receivers.

\subsection{Generality across HAR architectures}

\begin{figure}[!b]
  \centering
  \includegraphics[width=0.95\linewidth]{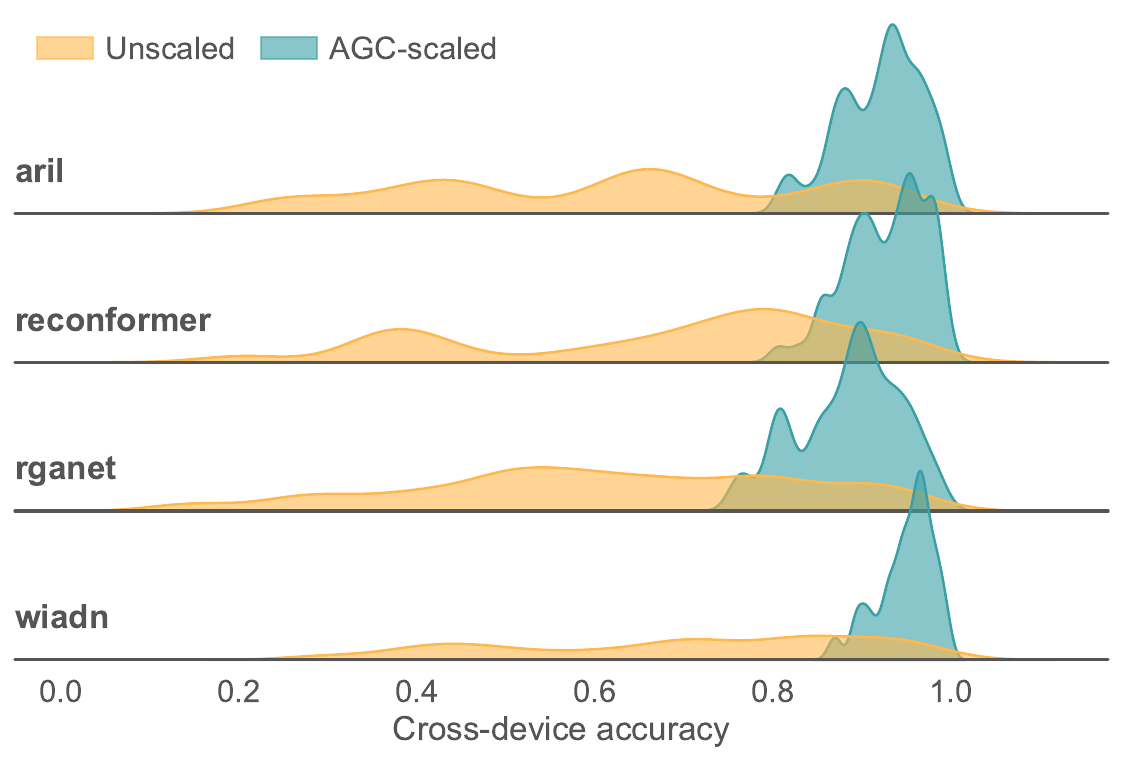}
  \caption{Distributions of cross-device accuracies for four \gls{har} models before and after gain normalization, using feature-wise z-score standardization across examples. Curves aggregate off-diagonal accuracies across all train--test device pairs.}
  \Description{Density plots of cross-device accuracies for ARIL, WiADN, a recurrent Conformer, and RGANet, comparing raw and gain-normalized inputs under feature-wise z-score standardization.}
  \label{fig:har-architectures-comparison}
\end{figure}

So far we have focused on ARIL. To test whether the same cross-device effects appear in other \gls{har} architectures, we also evaluate WiADN~\cite{wiadn}, a recurrent Conformer~\cite{reconformer}, and RGANet~\cite{rganet}. We follow the public code with only minimal changes required by our activity-only, $700$-sample setting: WiADN is run with activity loss only and a patched attention-mask generator that supports $700$-sample windows; the recurrent Conformer uses its UT-HAR configuration unchanged; and RGANet uses the released \texttt{UT\_HAR\_GRU\_Attention} and UT-HAR code path. We prepend the same feature-wise z-score standardization as in \Cref{subsec:cross-device} with no other architectural or hyperparameter changes.

\paragraph*{Results}
\Cref{fig:har-architectures-comparison} shows the resulting cross-device accuracy distributions. All four models suffer from clear cross-receiver degradation without gain normalization, despite standardization. After applying gain normalization, the distributions shift to higher accuracies for every architecture, and low-accuracy cases become rare.

\paragraph*{Take-away}
Cross-device sensitivity is not specific to ARIL: it appears across convolutional, recurrent, and attention-based \gls{har} models. Gain normalization, combined with simple feature-wise standardization, consistently improves cross-receiver compatibility for all architectures we test.

\begin{figure}[!t]
    \centering
    \includegraphics[width=0.95\linewidth]{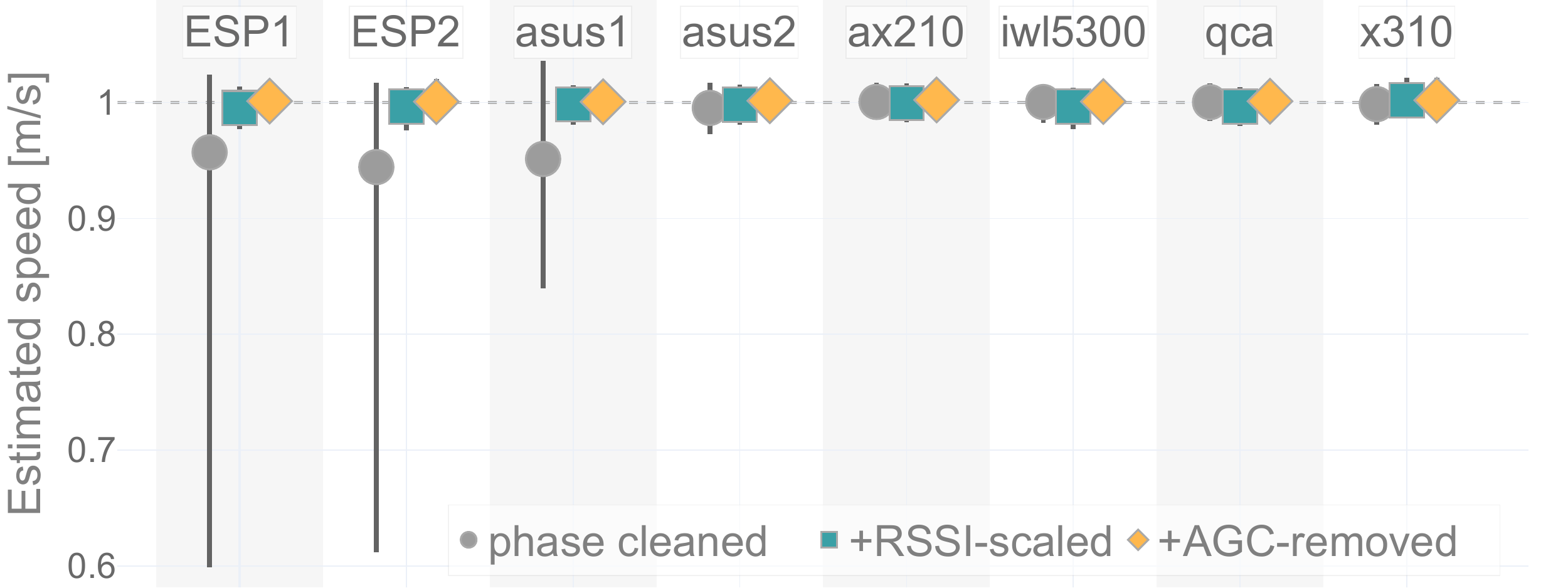}
    \caption{Median and \gls{iqr} of Doppler-MUSIC velocity estimates under the three preprocessing schemes; the dashed line marks the ground-truth $1\,\text{m/s}$.}
      \Description{Median and \gls{iqr} of Doppler-MUSIC velocity estimates under the three preprocessing schemes; the dashed line marks the ground-truth $1\,\text{m/s}$.}
    \label{fig:velocity-distribution}
\end{figure}

\subsection{Doppler-based velocity estimation}\label{subsec:doppler}

Having shown that large-scale gain confounds the \textit{learning} pipeline, we next test a \textit{purely signal-processing-based} method. We use Doppler-MUSIC~\cite{li2017indotrack,cominelli2024physical}, which estimates velocity from a \gls{csi} covariance matrix across packets and assumes that the signal statistics within each window are stable. Packet-to-packet gain changes from \gls{agc} break this assumption and should affect its estimates. To expose this effect clearly, we design a simple Doppler experiment with a single, perfectly known target velocity.

\paragraph*{Synthetic ground truth}
In real human motion, the measured Doppler is a superposition of reflections from multiple body parts. Even if a dominant component can often be isolated, its true velocity is not known, so we cannot quantify absolute error or bias against ground truth. We therefore emulate motion at the transmitter via packet-wise precoding $c_k(t) = a + b \cdot \exp\left(-j 2 \pi \; \frac{f_k \Delta(t)}{c} \right)$,
with $f_k$ the subcarrier frequency, $c$ the speed of light and $\Delta(t)=v t$ chosen to realize a constant velocity $v = 1\,\text{m/s}$.  
Receivers then observe a synthetic two-path channel
\begin{align*}
    \hat{H}_k(t) =
    \underbrace{a \cdot H_k^S}_{H_k^0} +
    \underbrace{b \cdot H_k^S \cdot \exp\left(-j 2 \pi \; \frac{f_k \Delta(t)}{c} \right)}_{H_k^1(t)},
\end{align*}
in which $a=0.7$ and $b=0.3$ fix the static-to-dynamic ratio.  The linear growth of $\Delta(t)$ gives a perfectly constant Doppler shift, so the true velocity is known exactly and we can directly measure estimator bias and spread.

\paragraph*{Data collection and processing}

We collect data at $500$ packets per second over $50$ seconds and apply Doppler-MUSIC to windows of $50$ \gls{csi} samples, using reported timestamps to estimate the effective gap time when packets are missing. For each packet, we remove a \gls{ls} linear phase fit across subcarriers to correct the linear terms in \Cref{eq:amp-phase-model}, which capture phase errors such as \gls{cfo}; \Cref{subsec:pads-phase-comp} considers an alternative using edge-subcarriers only to estimate slope. We always apply phase cleaning and then add either \gls{rssi}-based rescaling (\Cref{eq:rssi-rescale}) or large-scale-gain removal (\Cref{eq:amp-norm}).

\paragraph*{Results}

\begin{figure*}[!ht]
    \centering
    \subfloat[Raw \gls{csi} phase]{%
        \includegraphics[width=0.32\textwidth]{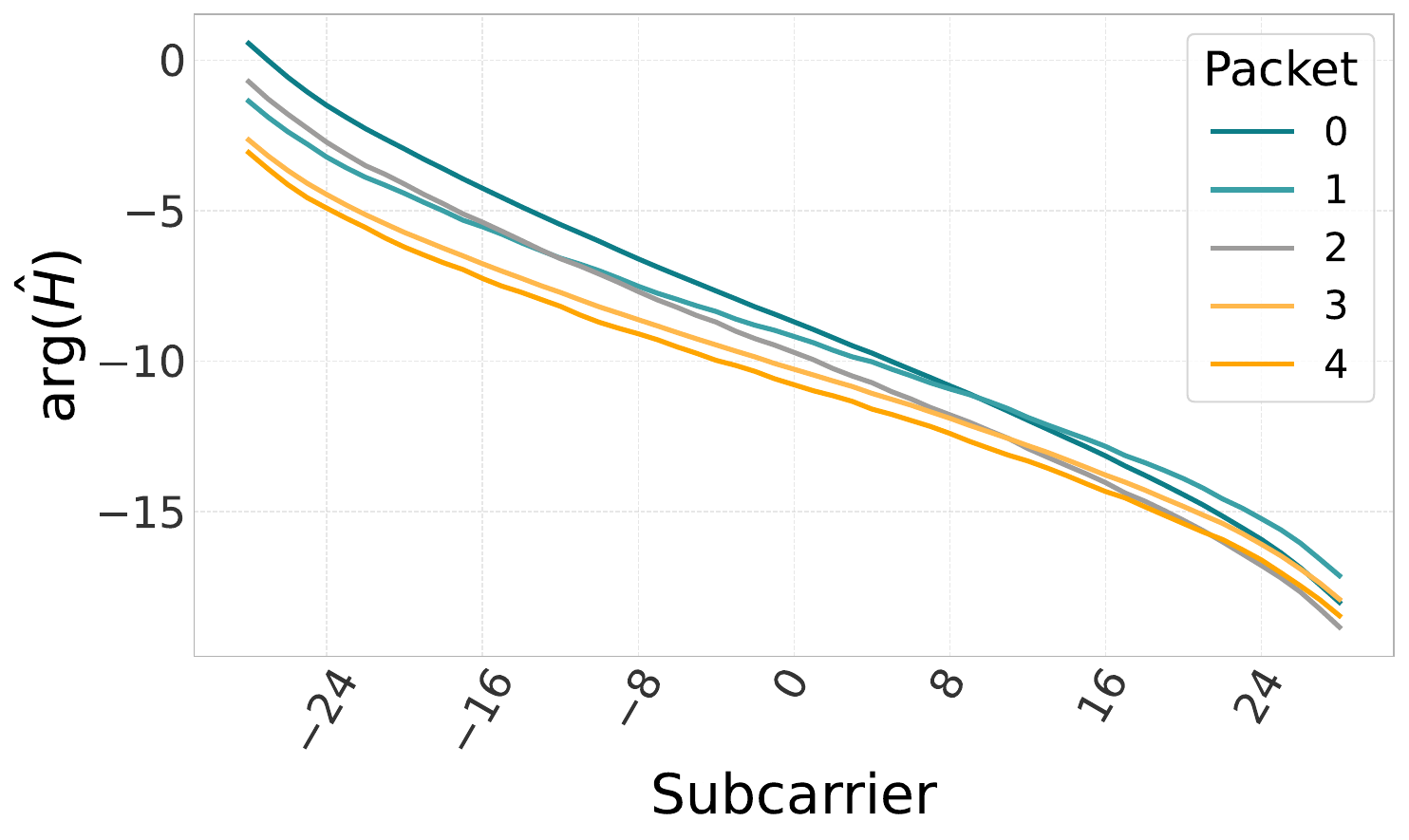}%
        \label{fig:raw-csi}%
    }\hfill
    \subfloat[After linear phase offset removal]{%
        \includegraphics[width=0.32\textwidth]{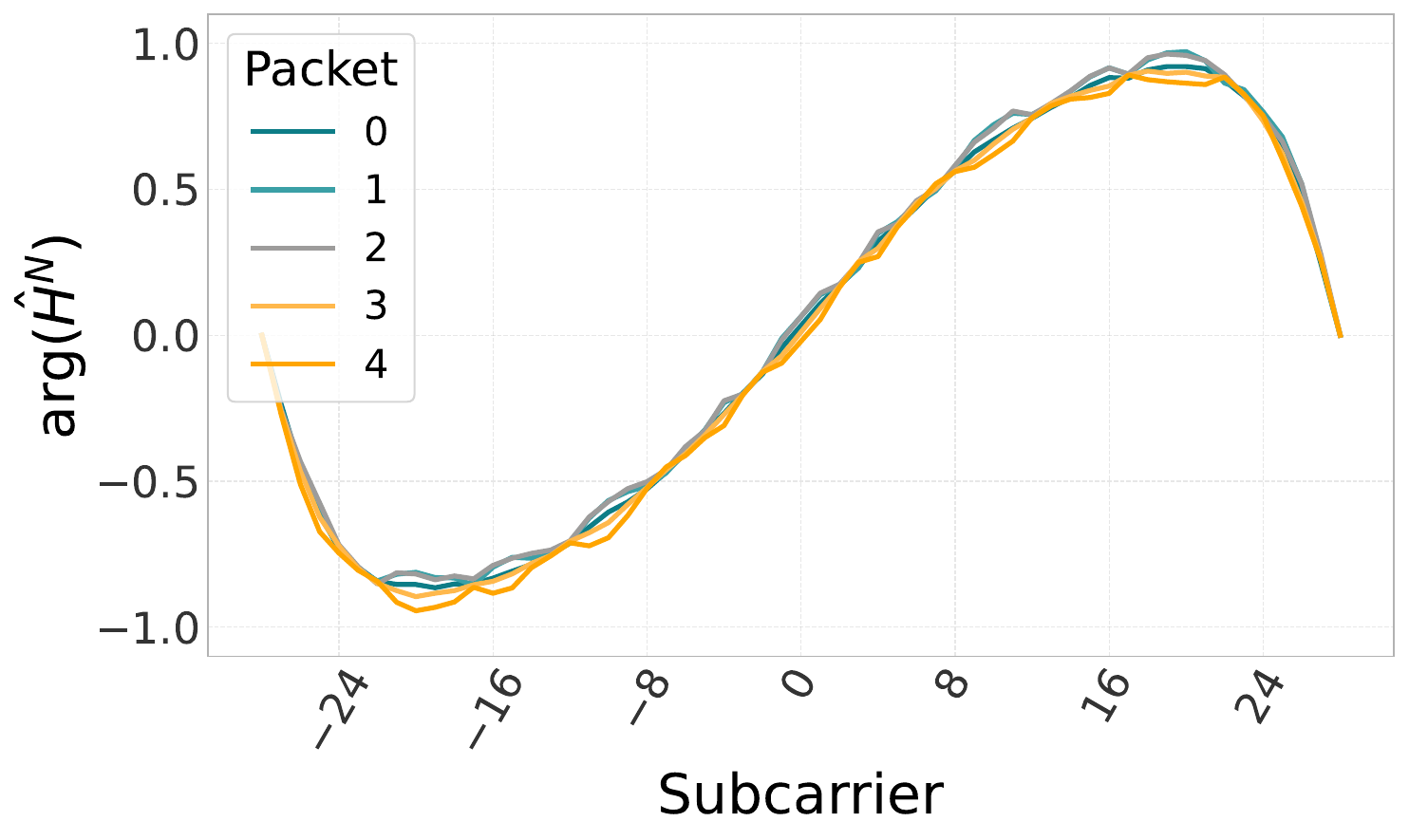}%
        \label{fig:normalized-csi}%
    }\hfill
    \subfloat[After consistent shape equalization]{%
        \includegraphics[width=0.32\textwidth]{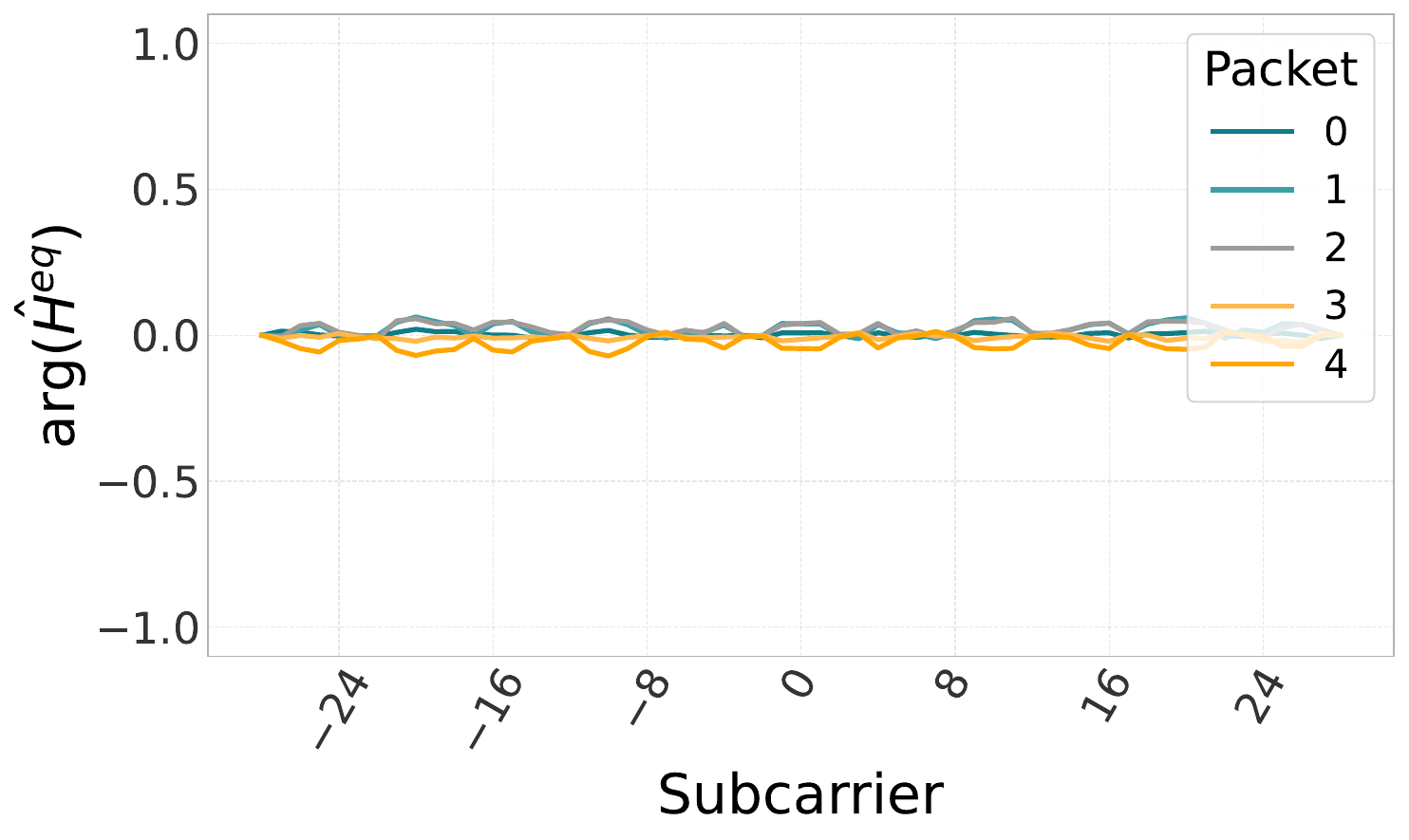}%
        \label{fig:equalized-csi}%
    }
    \caption{\gls{csi} preprocessing steps illustrated with \asustwo{}:
    \protect\subref{fig:raw-csi} raw \gls{csi} phase,
    \protect\subref{fig:normalized-csi} after linear phase offset removal,
    and \protect\subref{fig:equalized-csi} after consistent shape equalization.}
    \label{fig:csi-processing-stages}
    \Description{\gls{csi} preprocessing steps illustrated with \asustwo{}, showing raw phases, phases after linear detrending, and after shape equalization, revealing consistent device-specific distortions.}
\end{figure*}

\Cref{fig:velocity-distribution} shows that, with \gls{agc} intact, velocity estimates from the two \esp{} units and \asusone{} become \emph{highly unstable}: their \glspl{iqr} widen to up to $0.6\,\text{m/s}$, and their medians undershoot the $1\,\text{m/s}$ ground truth by $\approx 4\,\text{cm/s}$. The second \asus{} router (\asustwo{}) is almost unbiased, highlighting how device-internal \gls{agc} settings can \emph{differ even within a product line}. Applying \gls{rssi}-based rescaling (\Cref{eq:rssi-rescale}) \emph{collapses the spreads} and \emph{removes the large positive bias} in all three outliers, although a small negative offset persists in some medians. Eliminating large-scale gain \emph{cures both problems}: all medians now lie within $2\,\text{mm/s}$ of the target.

\paragraph*{Why time-invariant distortions are harmless to MUSIC}

The time-invariant distortions investigated in \Cref{sec:nonlinearity} are unimportant to this method. A deterministic, packet-constant distortion is a fixed diagonal weight $W$, yielding $Y = XW$. Its sample covariance is $R_Y = Y Y^{\mathrm H} = X (W W^{\mathrm H}) X^{\mathrm H}$, which rescales eigenvalues but leaves the signal and noise subspaces, and hence the MUSIC spectrum, unchanged. Doppler-MUSIC is therefore immune to any distortion that \emph{stays fixed across packets}.

\paragraph*{Take-away}
\gls{agc} makes Doppler-MUSIC velocity estimates highly unstable. \gls{rssi}-based rescaling removes most spread but leaves a \emph{small, systematic bias}, and is \emph{outperformed by large-scale-gain removal} that also rescued the ML pipeline. Large-scale gain is thus anything but irrelevant to Doppler-MUSIC; erratic gain control is harmful yet straightforward to eliminate.

\section{Nonlinear distortions} \label{sec:nonlinearity}

Beyond \gls{agc}-driven global amplitude scaling and random linear phase offsets, receiver hardware introduces subcarrier-dependent nonlinear distortions. As described in \Cref{subsec:csi-bg}, these effects, captured by the static terms $A_k$ and $\Phi_k$ in \Cref{eq:amp-phase-model}, persist even after correcting for global gain and linear phase offsets. Prior studies ($\pi$-Splicer~\cite{zhu2018pi}, PicoScenes~\cite{jiang2021eliminating}) observed these distortions on few devices, but did not systematically verify time invariance or impact on downstream sensing across diverse receiver types.

To characterize these subcarrier-specific nonlinearities, we collect \gls{csi} from repeated transmissions of an unmodified packet in a fully static, cable-connected setup (no precoding). For each receiver, we then apply three preprocessing steps (\Cref{fig:csi-processing-stages})

\paragraph*{Phase correction}
We unwrap phase across subcarriers and remove random linear phase offsets (see \autoref{eq:amp-phase-model}) by subtracting a line fitted between the first and last subcarriers, following PADS~\cite{qian2014pads}:
\begin{align*}
    a = \frac{\arg(\hat{H}_K) - \arg(\hat{H}_0)}{m_K - m_0}, \qquad b = \arg(\hat{H}_0),
\end{align*}
where $m_0$ and $m_K$ are indices of the first and last subcarriers. Using only the edge tones defines a fixed phase reference independent of the interior subcarriers, so that subcarrier-dependent distortions from precoding and hardware remain visible in the corrected \gls{csi}.

\paragraph*{Amplitude correction} Next, we remove \gls{agc}-induced amplitude scaling by normalizing each packet per \Cref{eq:amp-norm}. Remaining amplitude variations then reflect subcarrier-dependent nonlinearities on a shared scale.

\paragraph*{Profile computation} Finally, we compute amplitude and phase profiles by averaging the normalized, phase-corrected \gls{csi} over $N$ packets:
\begin{align}
    \overline{A}_k = \frac{1}{N} \sum_{n=1}^{N} |\hat{H}_{k,n}^N|, \qquad
    \overline{\Phi}_k = \frac{1}{N} \sum_{n=1}^{N} \arg(\hat{H}_{k,n}^N),
\end{align}
resulting in the baseline channel profile
\begin{align} \label{eq:csi-profile}
    \overline{H}_k = \overline{A}_k e^{j \overline{\Phi}_k}.
\end{align}

These static profiles $\overline{H}_k$ represent device-specific nonlinear distortions remaining after global corrections and serve as calibration references. In particular, we can equalize via
\begin{align} \label{eq:profile-equalization}
H^{eq} &= \frac{\hat{H}}{\overline{H}}.
\end{align}


\subsection{Consistency of distortions}

\begin{figure}[b]
    \centering
    \includegraphics[width=1.0\linewidth]{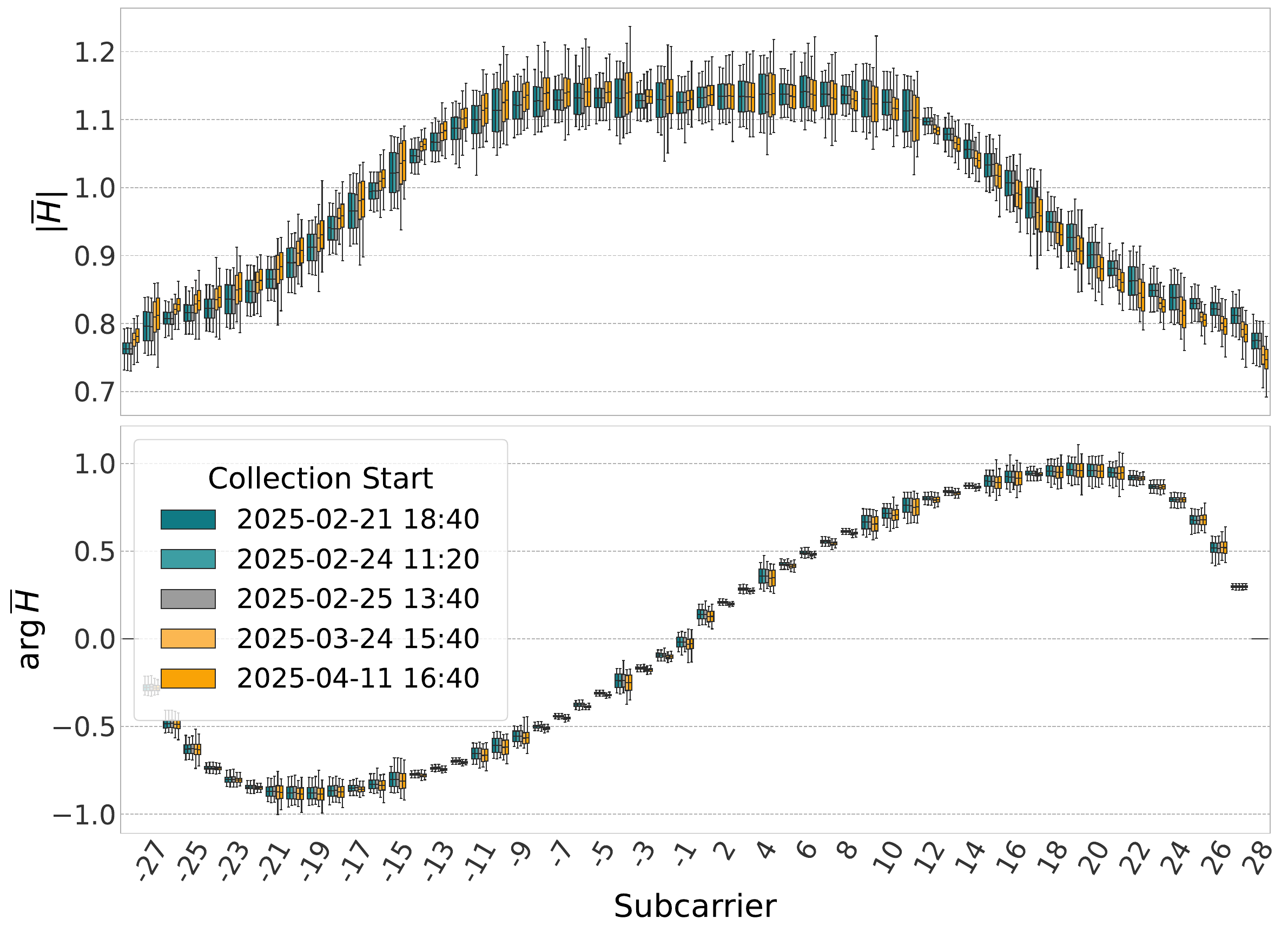}
    \caption{Nonlinear distortion boxplots for \asusone{} measured on five distinct days.}
    \Description{Nonlinear distortions for \asusone{} measured on five distinct days.}
    \label{fig:nonlinear-profiles-asus1}
\end{figure}
\begin{table}[!t]
    \centering
    \normalsize
    \caption{Stability scores for all receivers.}
    \begin{tabular}{p{5cm} c} 
        \toprule
        Receiver & Stability Score ($S$) \\
        \midrule
        x310 (SDR)         & 0.99998 \\
        ax210 (Intel)      & 0.99970 \\
        iwl5300 (Intel)    & 0.99966 \\
        qca (Qualcomm)     & 0.99997 \\
        asus1 (Broadcom)   & 0.99994 \\
        asus2 (Broadcom)   & 0.99997 \\
        ESP1 (Espressif)   & 0.99993 \\
        ESP2 (Espressif)   & 0.99993 \\
        \bottomrule
    \end{tabular}
    \label{table:stability-scores}
\end{table}

The extended distortion model in \Cref{eq:amp-phase-model} assumes that the device-specific nonlinear distortions $\overline{H}_k$ are time-invariant. We test this by repeating the profile extraction on five separate days under nominally identical laboratory conditions: on each day, we collect \gls{csi} from $10\,000$ repetitions of an unmodified packet, compute normalized channel profiles from \Cref{eq:csi-profile}, and then compare them across days.

\Cref{fig:nonlinear-profiles-asus1} illustrates this for \asusone{}: amplitude and phase medians, \glspl{iqr}, and extremes almost coincide across days, indicating highly stable distortions. For brevity, we omit the other devices and instead summarize stability using cosine similarities between profiles from days $i$ and $j$,
$S_{ij} = \frac{\langle \overline{H}^{(i)}, \overline{H}^{(j)} \rangle}{|\overline{H}^{(i)}||\overline{H}^{(j)}|}$,
where $\overline{H}^{(i)}$ is the day-$i$ profile. For each device we report the geometric mean
$S = \left( \prod_{i<j} S_{ij} \right)^{2 / (n(n-1))}$ over all day pairs, with $n$ the number of measurement days. This score equals one only if all profiles are identical, and values close to one reflect strong day-to-day consistency. Stability scores for all receivers are given in \Cref{table:stability-scores}.

All devices achieve high stability scores, showing that the subcarrier-dependent nonlinear distortions are effectively static over days. This supports the static-distortion assumption in \Cref{eq:amp-phase-model} and justifies treating the extracted profiles as fixed calibration terms in subsequent experiments.


\begin{figure}[!b]
    \centering
    \includegraphics[width=0.95\linewidth]{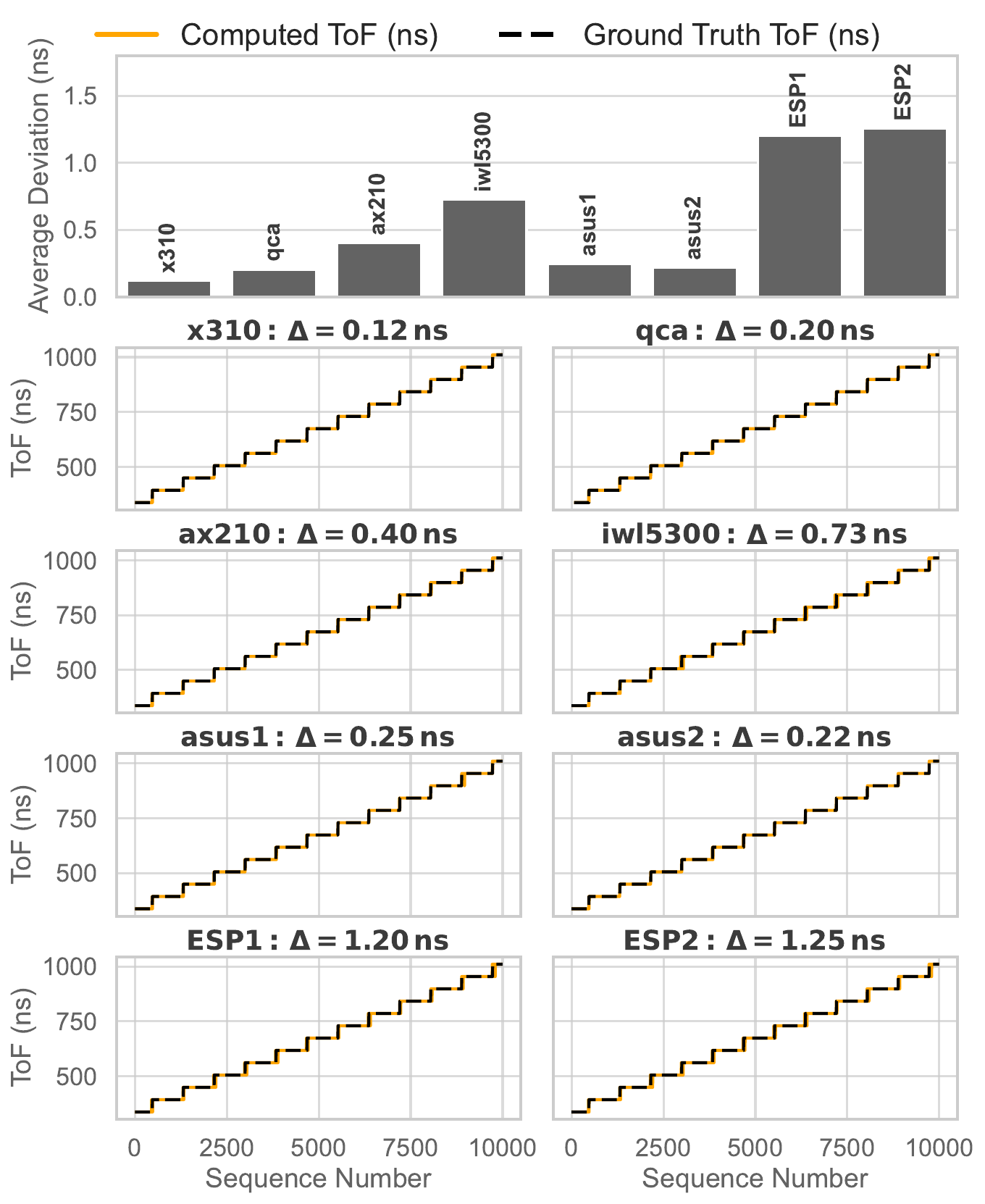}
    \caption{Time of flight estimates from \gls{pdp} and ground truth over time.}
    \Description{Time of flight estimates from \gls{pdp} and ground truth over time.}
    \label{fig:pdp-tof}
\end{figure}

\begin{figure}[!t]
    \centering
    \includegraphics[width=0.9\linewidth]{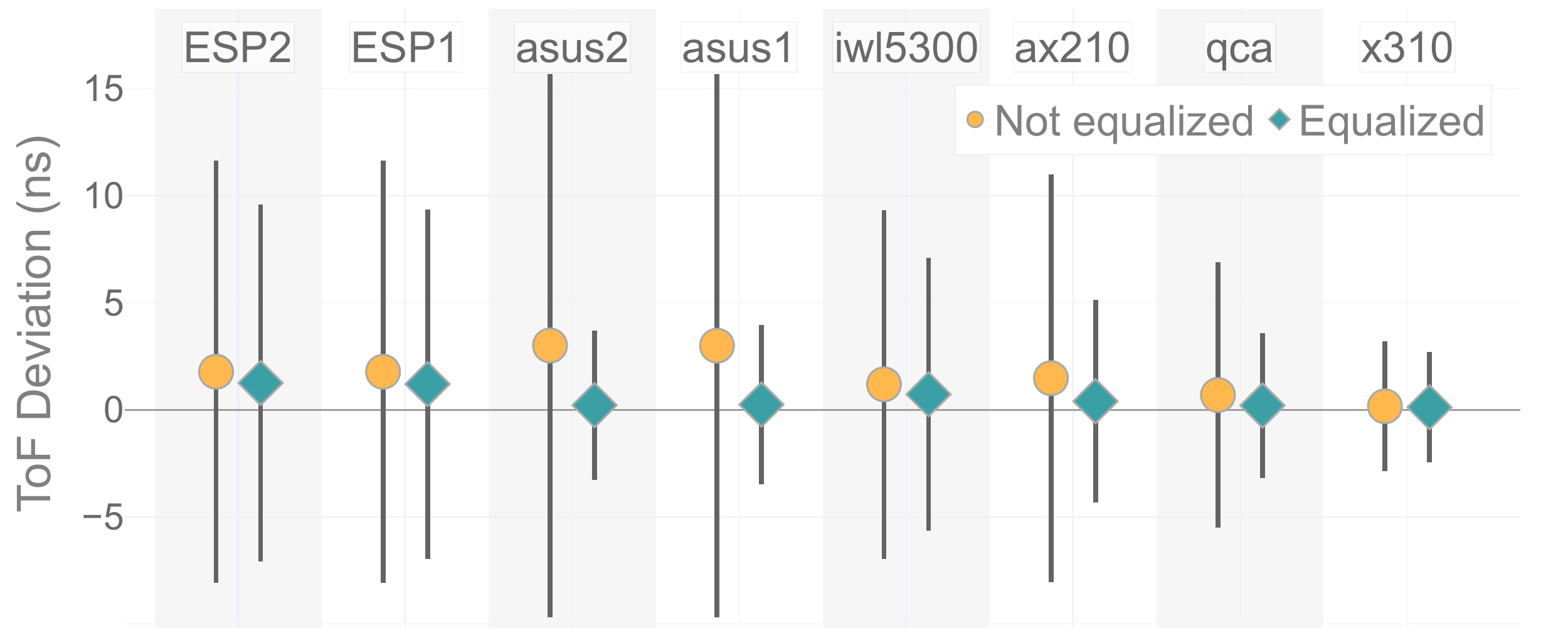}
    \caption{\gls{tof} deviation from ground truth with and without nonlinearity equalization.}
    \Description{\gls{tof} deviation from ground truth with and without nonlinearity equalization.}
    \label{fig:tof-equalization-improvement}
\end{figure}

\begin{figure}[!t]
    \centering
    \includegraphics[width=0.7\linewidth]{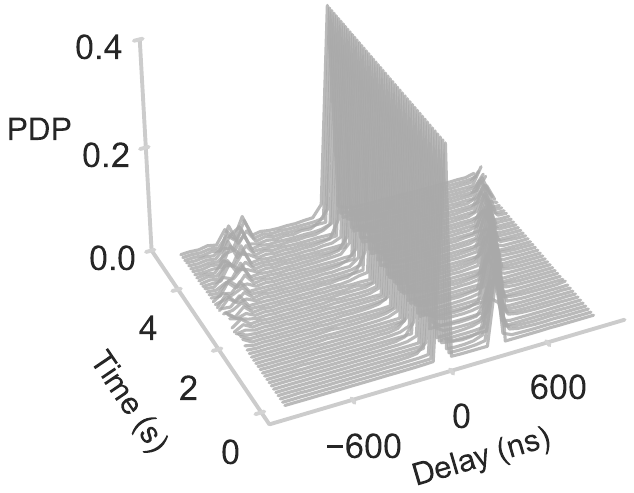}
    \caption{\gls{pdp} of \iwl{} over time; truncated to $0.4$ for visualization purposes.}
    \Description{\gls{pdp} of \iwl{} over time; truncated to $0.4$ for visualization purposes.}
    \label{fig:pdp-waterfall-intel}
\end{figure}

\subsection{PDP-based ToF estimation} \label{subsec:pdp-tof}

Subcarrier nonlinearities also affect \gls{pdp}-based \gls{tof} estimation. We compute the \gls{pdp} as $\text{PDP}(\tau) = \left|\mathcal{F}^{-1}\{\hat{H}\}(\tau)\right|^2$, and again remove linear phase offsets with a \gls{ls} fit (see \Cref{subsec:pads-phase-comp} for a comparison with PADS). Because the DC subcarrier is absent in \gls{csi} while the \gls{fft} requires equally spaced frequency samples, we reconstruct the DC term by linear interpolation and then estimate \gls{tof} from prominent \gls{pdp} peaks~\cite{xie2015precise}. Specifically, we detect the two strongest peaks, take their positive delay difference as the path delay, and tune the peak-detection threshold separately for each device.

For accurate evaluation, we reuse the precoding scheme from \Cref{subsec:doppler}, now letting $\Delta(t)$ increase linearly from $100$ to $300$\,m over $50$\,s to induce a known \gls{tof} variation. We measure performance as the deviation between estimated \gls{tof} and ground truth (\Cref{fig:pdp-tof}). Because each device’s bandwidth $B$ limits its delay resolution to steps of $1/B$, estimates show a step-like behavior.

Nonlinear distortions substantially reshape the \gls{pdp}. \Cref{fig:tof-equalization-improvement} shows that equalizing them with \Cref{eq:profile-equalization} clearly improves mean \gls{tof} accuracy and volatility across all devices. Among the tested receivers, \usrp{} and \qca{} achieve the best accuracy, followed by Intel \glspl{nic}, while \asus{} routers gain the most from equalization.

To isolate the effect of reduced effective bandwidth, we show the \gls{pdp} for Intel \iwl{} in \Cref{fig:pdp-waterfall-intel}; other devices do not exhibit this artifact. \iwl{} reports \gls{csi} for only $30$ subcarriers, effectively doubling the subcarrier spacing and introducing aliasing near $1600$\,ns. This yields prominent negative-delay peaks (a \gls{tof} ambiguity artifact) and degrades peak resolution at higher delays.

\begin{figure*}[!t]
    \centering
    \includegraphics[width=0.9\textwidth,alt={Noise correlation between subcarrier for all receivers.}]{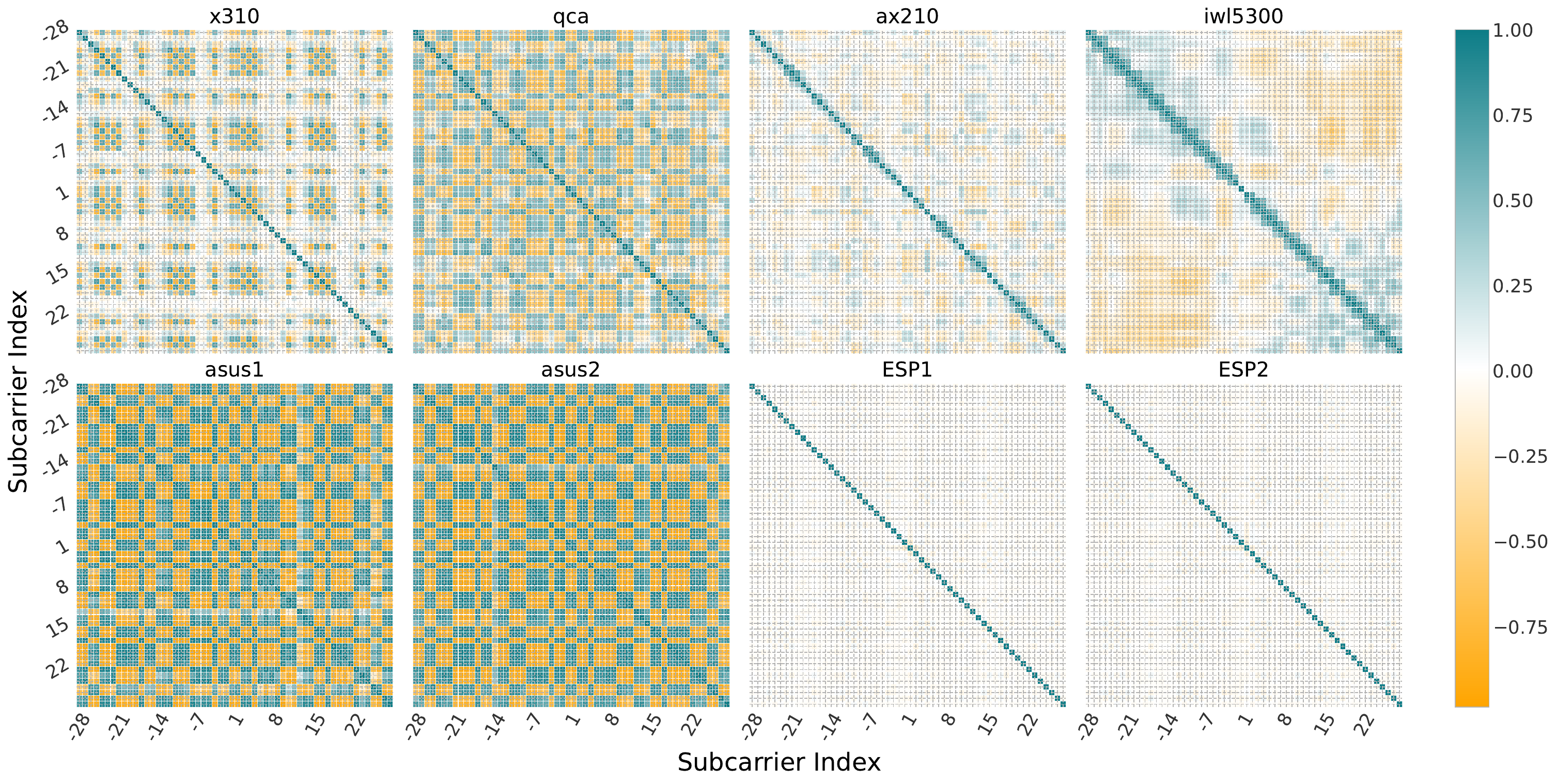}
    \caption{Static-environment correlation of $\hat{H}'$ between different subcarriers for all receivers.}
    \Description{Static-environment correlation of $\hat{H}$ between different subcarriers for all eight receivers clearly showing filtering effects.}
    \label{fig:noise-correlation}
\end{figure*}

\section{Beyond tractable distortions} \label{sec:beyond}

Our focus so far has been on distortions that are, at least in theory, predictable and correctable. However, \gls{csi} measurements are also inherently noisy, and their accuracy is limited by hardware-specific constraints and imperfections in the estimation algorithm.

We now probe noise, faithfulness, and sensitivity of the estimator $E$, beyond \Cref{eq:amp-phase-model}. We selectively precode a subset of subcarriers to isolate localized effects against a shared baseline. Preprocessing follows \Cref{sec:nonlinearity} except that precoded tones are excluded from the gain calculation, anchoring unmodified tones. Concretely, $\hat{H}^N_k = \frac{\hat{H}_k}{\sum{u \in U} |\hat{H}_u|}$,
where $U$ denotes the set of non-precoded subcarriers. This removes the effects in \Cref{eq:amp-phase-model}; remaining differences between precoded and unmodified tones reflect estimation error or residual, non-modeled distortions.


\subsection{Noise characteristics} \label{sec:noise}

\begin{table}[!b]
    \centering
    \normalsize
    \caption{Amplitude noise metrics for different receivers, summarizing \gls{std}, \gls{iqr}, skewness, and kurtosis.}
    \begin{tabular}{lcccc}
        \toprule
        Receiver & \gls{std} & \gls{iqr} & Skewness & Kurtosis \\
        \midrule
        x310 (SDR)         & 0.003 & 0.004 & 0.006   & 0.148    \\
        ax210 (Intel)      & 0.004 & 0.006 & 0.147   & 0.444    \\
        iwl5300 (Intel)    & 0.013 & 0.018 & -0.049  & -0.277   \\
        qca (Qualcomm)     & 0.015 & 0.023 & 0.005   & -0.525   \\
        asus1 (Broadcom)   & 0.015 & 0.026 & -0.450  & 43.259   \\
        asus2 (Broadcom)   & 0.017 & 0.031 & -0.285  & 17.710   \\
        ESP1 (Espressif)   & 0.035 & 0.048 & -0.005  & -0.016   \\
        ESP2 (Espressif)   & 0.037 & 0.050 & 0.003   & -0.002   \\
        \bottomrule
    \end{tabular}
    \label{tab:magnitude_noise}
\end{table}

\begin{table}[!b]
    \centering
    \normalsize
    \caption{Phase noise metrics for different receivers, summarizing \gls{std}, \gls{iqr}, skewness, and kurtosis.}
    \begin{tabular}{lcccc}
        \toprule
        Receiver & \gls{std} & \gls{iqr} & Skewness & Kurtosis \\
        \midrule
        x310 (SDR)         & 0.004 & 0.005 & -0.013  & 0.405    \\
        ax210 (Intel)      & 0.007 & 0.009 & -0.046  & 0.117    \\
        iwl5300 (Intel)    & 0.018 & 0.025 & -0.030  & -0.203   \\
        qca (Qualcomm)     & 0.018 & 0.028 & 0.110   & 0.219    \\
        asus1 (Broadcom)   & 0.020 & 0.032 & 1.968   & 87.702   \\
        asus2 (Broadcom)   & 0.020 & 0.036 & 2.832   & 176.879  \\
        ESP1 (Espressif)   & 0.049 & 0.066 & 0.002   & 0.319    \\
        ESP2 (Espressif)   & 0.052 & 0.070 & 0.002   & 0.288    \\
        \bottomrule
    \end{tabular}
    \label{tab:phase_noise}
\end{table}

We analyze equalized \gls{csi} $H^{eq}$ from $10\,000$ repetitions of an unmodified packet under controlled conditions. We chose per-subcarrier metrics \gls{std}, \gls{iqr}, skewness, and excess kurtosis. Because these ignore cross-tone dependencies, we also quantify cross-subcarrier noise correlations (sometimes assumed zero; \eg \cite{wang2016csi,vogt2019precise}), which may accumulate when averaging across tones and have not been systematically examined.

\subsubsection{Amplitude Noise}

We analyze amplitude noise by computing the above metrics per subcarrier and then averaging across subcarriers. \Cref{tab:magnitude_noise} summarizes results per receiver. The \usrp{} (SDR) shows the lowest noise (lowest \gls{std} and \gls{iqr}). Among \gls{cots} devices, the \intelax{} (Intel) stands out with a particularly low \gls{std}, likely benefiting from internal subcarrier smoothing (see \Cref{sec:ax210smoothing}). The \iwl{} (Intel), \qca{} (Qualcomm), and \esp{} (Espressif) exhibit near-normal amplitude noise (skewness and excess kurtosis close to zero). In contrast, \asusone{} and \asustwo{} (Broadcom) show extremely high kurtosis ($>40$), not as precise values but as indicators of rare, severe outliers; for example, removing around $20$ Mahalanobis-distance outliers on \asusone{} reduces kurtosis from $43.259$ to $0.82$, with a similar improvement for \asustwo{}. This underscores the value of reporting skewness and kurtosis alongside \gls{std} and \gls{iqr}, which can hide such outlier behavior.


\subsubsection{Phase Noise}

We perform the same analysis for phase noise. \Cref{tab:phase_noise} summarizes metrics per receiver. Consistent with the amplitude results, the \usrp{} shows the lowest variability, followed by the \intelax{}. The \iwl{} (Intel), \qca{} (Qualcomm), and \esp{} (Espressif) exhibit near-normal phase noise (skewness and kurtosis close to zero), whereas \asusone{} and \asustwo{} (Broadcom) show extreme phase kurtosis, indicating rare but large outliers. When outliers identified by \emph{amplitude} deviations are removed, both skewness and kurtosis return to levels comparable to the other \glspl{nic}, corroborating that these samples are invalid \gls{csi}.


\subsubsection{Noise Correlation}

We further examine noise correlation across subcarriers. \Cref{fig:noise-correlation} shows that \esp{} microcontrollers, despite higher per-subcarrier noise power, exhibit minimal cross-subcarrier correlation. In contrast, the \asus{} routers and the \qca{} \gls{nic} show strong cross-subcarrier correlations with near-identical patterns, suggesting a shared hardware component (\eg a filtering stage).

Overall, while receivers differ in noise characteristics, all maintain \emph{low} absolute noise. Cross-subcarrier correlations, particularly in the \asus{} routers, introduce an additional factor that \emph{may influence downstream performance}. \emph{High-fidelity} applications may favor the \usrp{}, and users of \asus{} devices should consider mitigation strategies for rare strong outliers. Ultimately, receiver choice depends on an application’s tolerance for noise magnitude, correlation, and outlier events.

\subsection{Faithfulness} \label{subsec:faithfulness}

Accurate channel estimation is \emph{fundamental} for the reliability of \gls{csi}-based applications. A \emph{strong alignment} between the estimated channel and the actual channel ensures that signal processing methods, which often rely on \emph{precise channel models}, function as intended. \emph{Discrepancies} between the estimated and true channel can \emph{violate these core assumptions}, undermining the effectiveness of downstream applications.

To quantify this alignment, we introduce the concept of \textbf{faithfulness}, which measures how closely the channel estimate mirrors the actual channel characteristics. Specifically, we define the \textbf{mean phase and amplitude response deviations} as follows:
\begin{align*}
    \bar{D}^{amp}_k(H) &= \frac{1}{N} \sum_{n=1}^N \left| |H^{eq}_{k,n}| - |c_n| \right|, \\
    \bar{D}^{phs}_k(H) &= \frac{1}{N} \sum_{n=1}^N \left| \arg(H^{eq}_{k,n}) - arg(c_n) \right|,
\end{align*}
where $c_n$ is the introduced complex precoding factor in the $n$-th out of a total of $N$ packets and $\hat{H}'_{n,k}$ the corresponding detected and equalized \gls{csi} on subcarrier $k$.

We use $N=1\,000$ packets and vary precoding amplitudes equidistantly between $[0,2]$, and phases between $[0, \tfrac{\pi}{2}]$.

\begin{figure}[!ht]
    \centering
    \subfloat[Mean amplitude response deviation. \label{fig:amp-meandev}]{
        \includegraphics[width=0.95\linewidth]{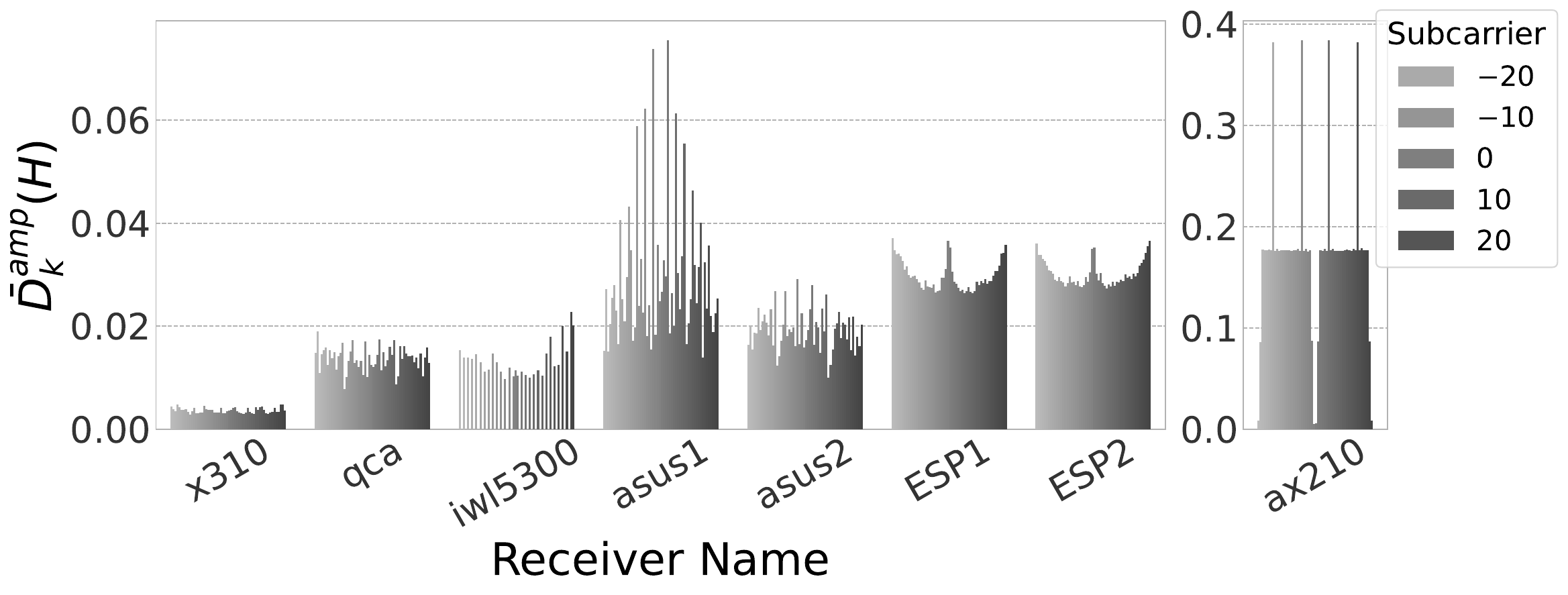}
        \Description{Mean amplitude response deviation.}
    }

    \subfloat[Mean phase response deviation \label{fig:phs-meandev}]{
        \includegraphics[width=0.95\linewidth]{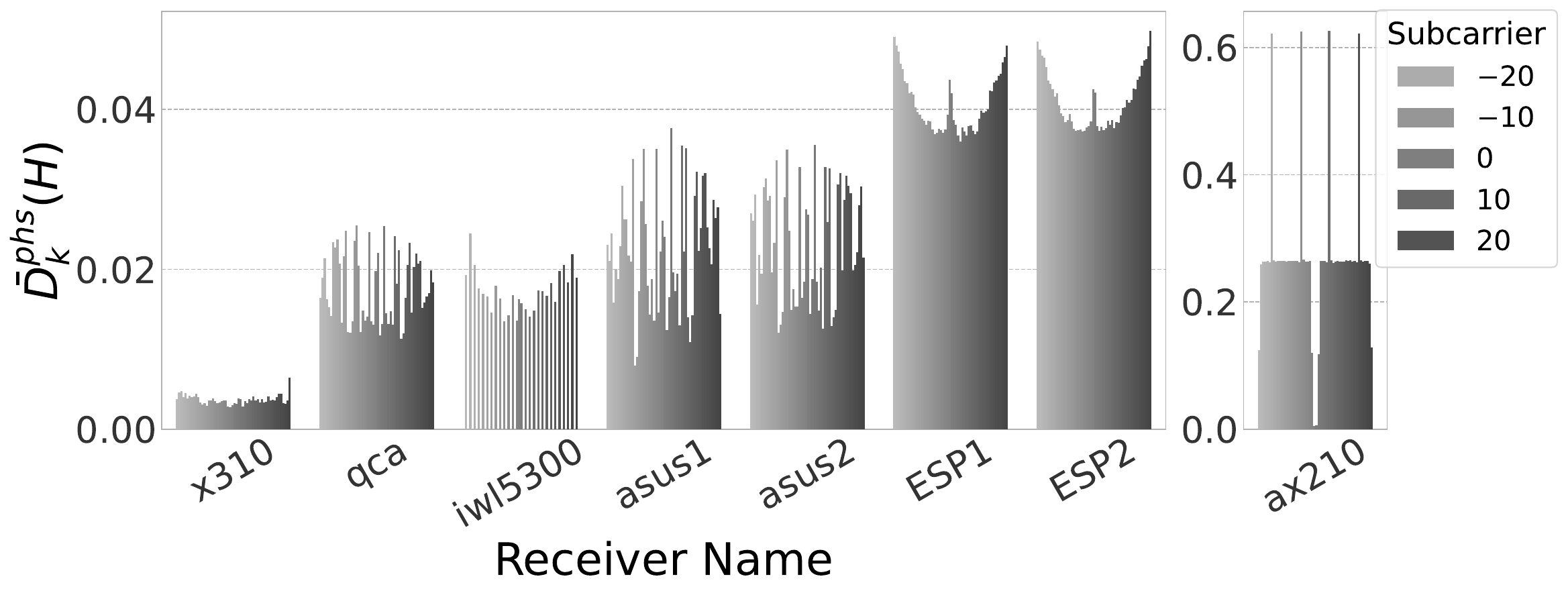}
        \Description{Mean phase response deviation.}
    }

    \caption{Mean response deviations for all devices and subcarriers.}
    \label{fig:all-meandev}
    \Description{Mean response deviations for all devices and subcarriers.}
\end{figure}

The results in \Cref{fig:amp-meandev} and \Cref{fig:phs-meandev} highlight notable differences in faithfulness across devices. With the exception of the \intelax{}, all devices demonstrate high faithfulness, albeit with some variation. As expected due to its premium build quality, the \usrp{} boasts both the least overall deviations as well as the highest consistency across subcarriers. For the others, a pronounced subcarrier dependency is evident. \asus{} devices and \qca{} exhibit larger deviations on a non-systematic set of subcarriers, especially pronounced in phase. The \esp{}s, on the other hand, show a clear shape, with slightly larger deviations towards edge and DC subcarriers. Notably, there is a strong difference between the two \asus{} routers, confirming that even devices of the same family can vary strongly.
The \iwl{} performs remarkably well, given its age, showcasing only a slight dependence on frequency and a comparatively low mean phase response deviation. The \intelax{} stands out as an outlier, exhibiting significant deviations in both amplitude and phase. We examine these anomalies further in \Cref{sec:ax210smoothing}.

Notably, the order of average deviation closely aligns with the ranking in accuracies observed in \Cref{subsec:pdp-tof}, confirming that \emph{device capabilities are distinct in high-precision tasks}.

\subsection{Intel ax210 smoothing} \label{sec:ax210smoothing}

The large deviation from the expected scaling on individual subcarriers observed in the Intel \intelax{} \gls{csi} measurements stems from a smoothing mechanism implemented by the device. To demonstrate this, we apply a magnitude precoding factor of $2$ across a larger block of subcarriers.

\begin{figure}[!b]
    \centering
    \includegraphics[width=1.0\linewidth]{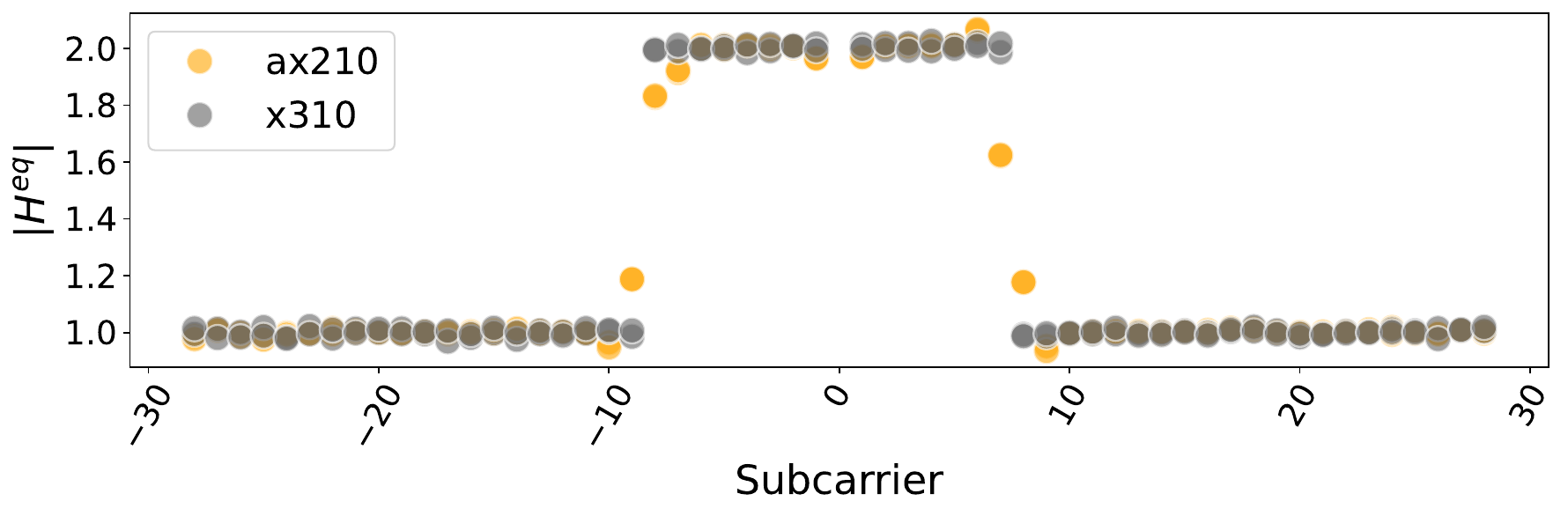}
    \caption{Detected magnitude scaling after precoding a block of subcarriers by a factor of $2$.}
    \Description{Detected magnitude scaling after precoding a block of subcarriers by a factor of $2$ showing an effect beyond modified subcarriers.}
    \label{fig:ax210-averaging}
\end{figure}

\Cref{fig:ax210-averaging} compares the reported \gls{csi} from the \intelax{} to that of the \usrp{}. In the center of the block, both devices detect the correct scaling. However, the \intelax{} exhibits a consistent smoothing effect at the block edges, suggesting an internal averaging process. Notably, this smoothing also affects phase measurements, even though phase values remain unaltered by the applied precoding. A similar smoothing effect is observed when applying phase precoding across a block of subcarriers.

While this may be relevant for applications with a narrow coherence bandwidth, we could not observe any negative impact in any of our tests, as demonstrated in the presented applications. Neither \gls{pdp}, Doppler velocity, nor neural networks manifest in a measurable negative performance hit for the \intelax{}.

\subsection{Sensitivity} \label{subsec:sensitivity}

\begin{figure}[!b]
    \centering
    \subfloat[Average amplitude sensitivity. \label{fig:amp-sensitivity-scores}]{
        \includegraphics[width=0.95\linewidth]{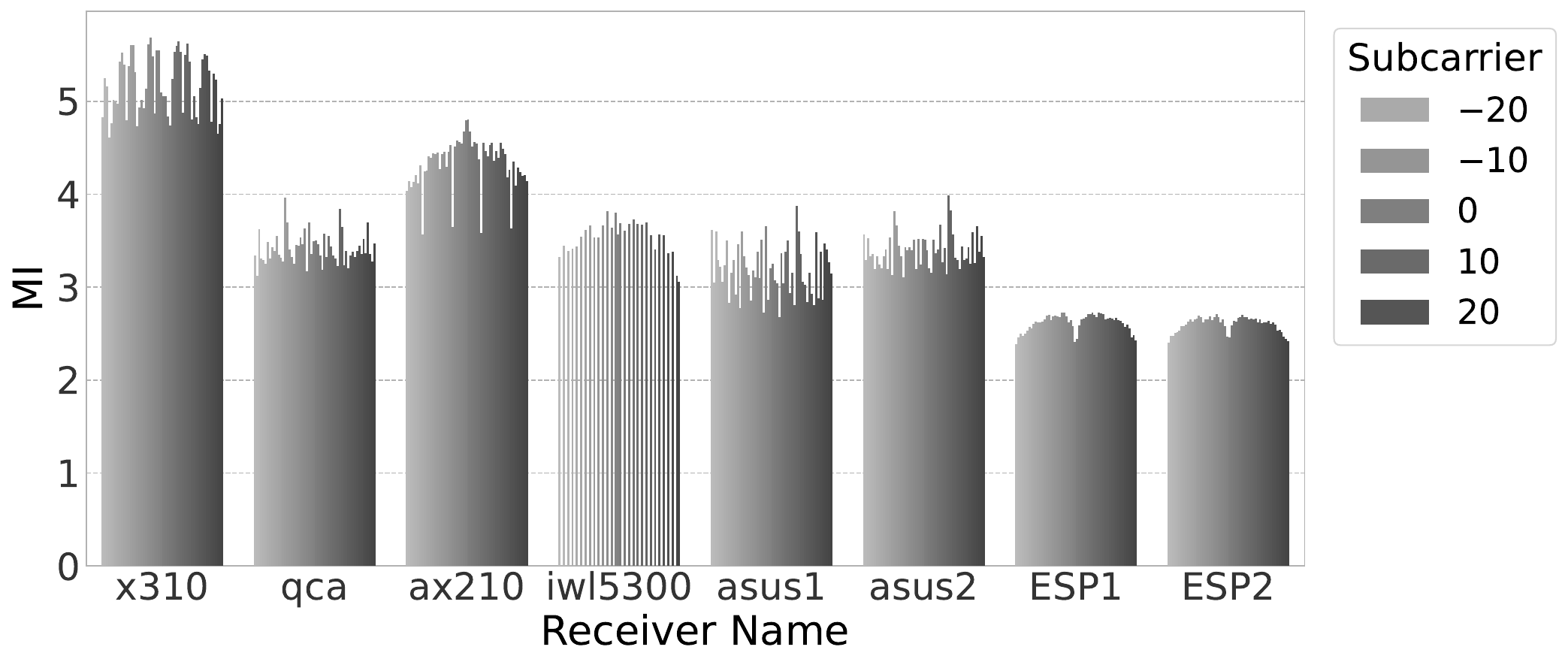}
        \Description{Average amplitude sensitivity for all devices and subcarriers, quantified with mutual information. Larger values (more bits) indicate stronger coupling between channel variations and reported estimates.}
    }

    \subfloat[Average phase sensitivity. \label{fig:phs-sensitivity-scores}]{
        \includegraphics[width=0.95\linewidth]{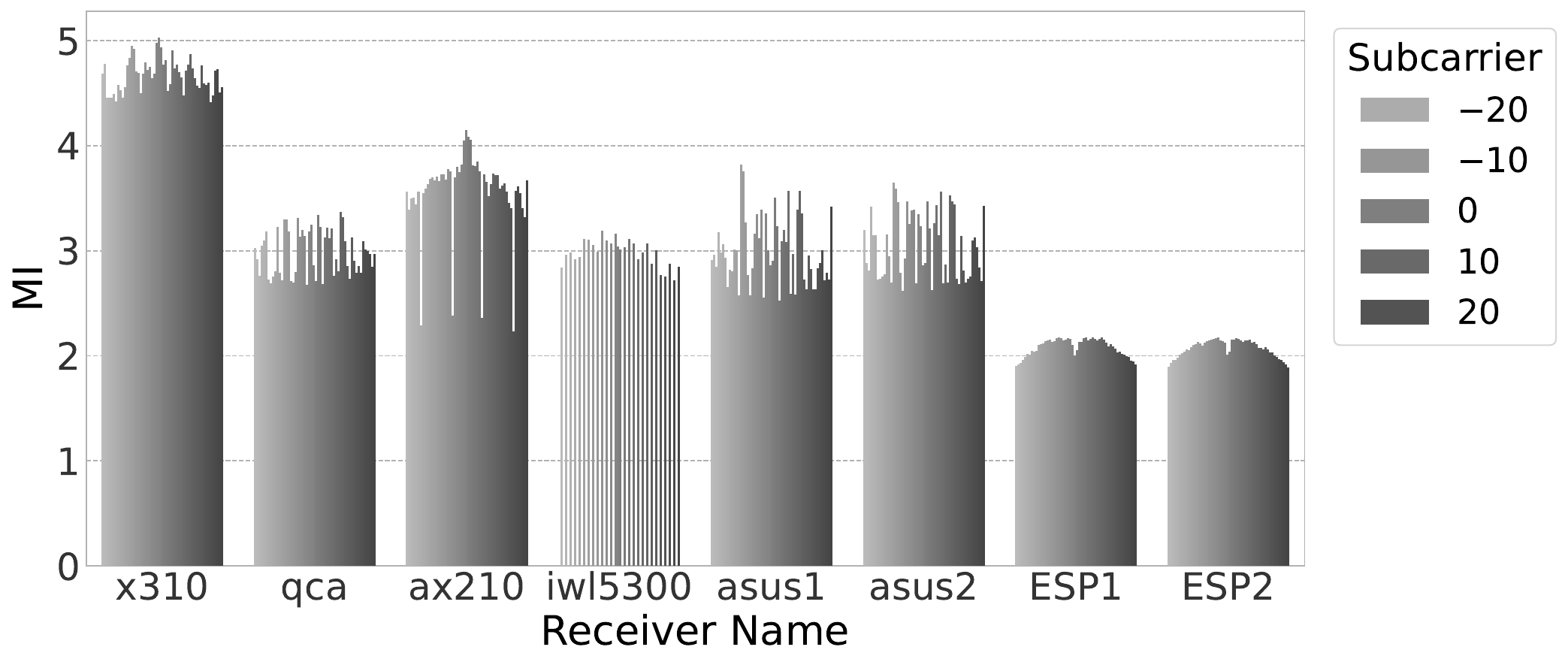}
        \Description{Average phase sensitivity for all devices and subcarriers, quantified with mutual information. Larger values (more bits) indicate stronger coupling between channel variations and reported estimates.}
    }

    \caption{Average sensitivity for all devices and subcarriers, quantified with mutual information. Larger values indicate stronger coupling between channel variations and reported estimates.}
    \label{fig:all-sensitivity}
    \Description{Average sensitivity for all devices and subcarriers, quantified with mutual information. Larger values indicate stronger coupling between channel variations and reported estimates.}
\end{figure}

Neural networks have changed the rules for \gls{csi}-based applications: \emph{fidelity no longer defines data quality}; \emph{exploitable structure does}. What matters is whether the data moves with the channel in a \emph{consistent, learnable pattern}. If it does, neural networks can extract meaningful insights. If it does not, \emph{no procedure, no matter how powerful, can uncover what is not there}.

To reveal how much exploitable structure each device provides, we measure the \emph{sensitivity} to the controlled channel variations $c_k$ in the reported estimates $\hat{H}_k'$ with \gls{mi}, $I(c_k;\hat{H}_k')$. \gls{mi} answers the precise question we care about: \emph{how much information about the channel does the \gls{csi} reveal?} Formally,

\[
I(c_k;\hat{H}_k') =
\iint
  p_{c,\hat{H}'}(c_k,\hat{H}_k')\,
  \ln\!\frac{p_{c,\hat{H}'}(c_k,\hat{H}_k')}{p_c(c_k)\,p_{\hat{H}'}(\hat{H}_k')}
  \,\mathrm{d}c_k\,\mathrm{d}\hat{H}_k',
\]

where $p_{c,\hat{H}'}$ is the joint density and $p_c$, $p_{\hat{H}'}$ the marginals. The logarithm is natural, so $I$ is in \emph{nats}. A value of $I=0$ implies statistical independence; larger values indicate stronger dependence and greater potential for downstream procedures to distinguish channel variations.

We estimate $I$ with the $k$-nearest-neighbor method of Kraskov~\cite{kraskov2004estimating}.

\paragraph*{Results}
The \gls{mi} mirrors the faithfulness analysis in \Cref{subsec:faithfulness} (\Cref{fig:all-sensitivity}). We again see the \usrp{} leading by a significant margin. The \intelax{}’s averaging is irrelevant for \gls{mi}; it shows the second-highest values across subcarriers, with slight dips on pilots. \qca{} and \iwl{} perform similarly, strong despite their age, and on levels comparable to the \asus{} routers. While \asusone{} appears slightly worse than its twin, the difference does not seem significant. The \esp{}s bring up the rear, behaving similarly and showing a small nudge around DC.

All devices yield \gls{mi} well above zero, confirming that each subcarrier conveys substantial information about the underlying channel. Absolute gaps are relatively small, suggesting that despite architectural differences, every device captures a similarly large amount of exploitable structure. We therefore expect similar performance on coarse-grained tasks, with \emph{subtle \gls{mi} differences} influencing only applications demanding high precision, as in \Cref{subsec:pdp-tof}.

\subsection{Channel, band and antenna dependence}

\begin{figure}[!b]
    \centering
    \subfloat[Amplitude response deviation. Lower is better. \label{fig:channel-dep-abs-prd}]{
        \includegraphics[width=0.95\linewidth]{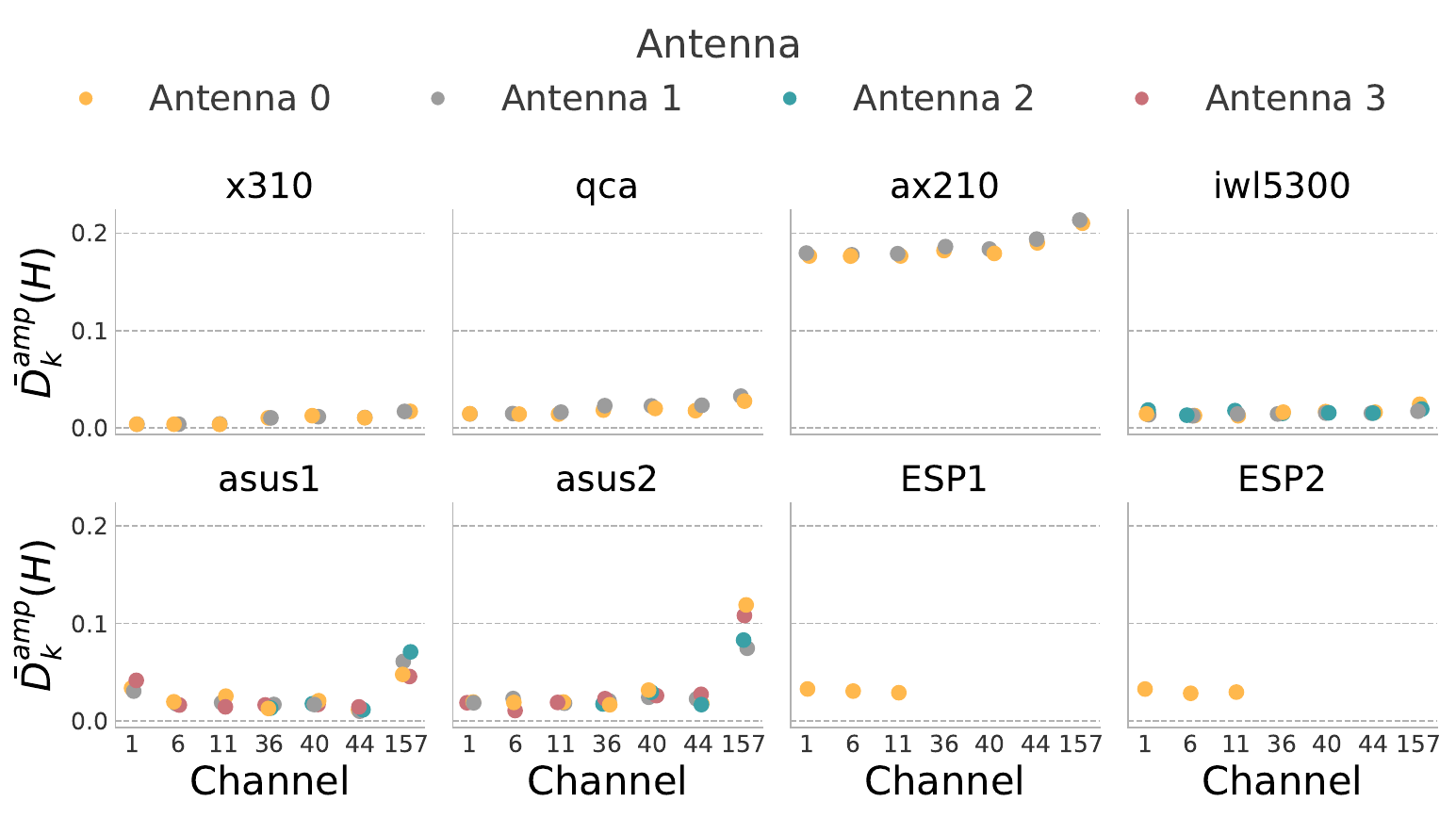}
        \Description{Amplitude response deviation. Lower is better.}
    }

    \subfloat[Amplitude sensitivity. Higher is better. \label{fig:channel-dep-abs-sensitivity}]{
        \includegraphics[width=0.95\linewidth]{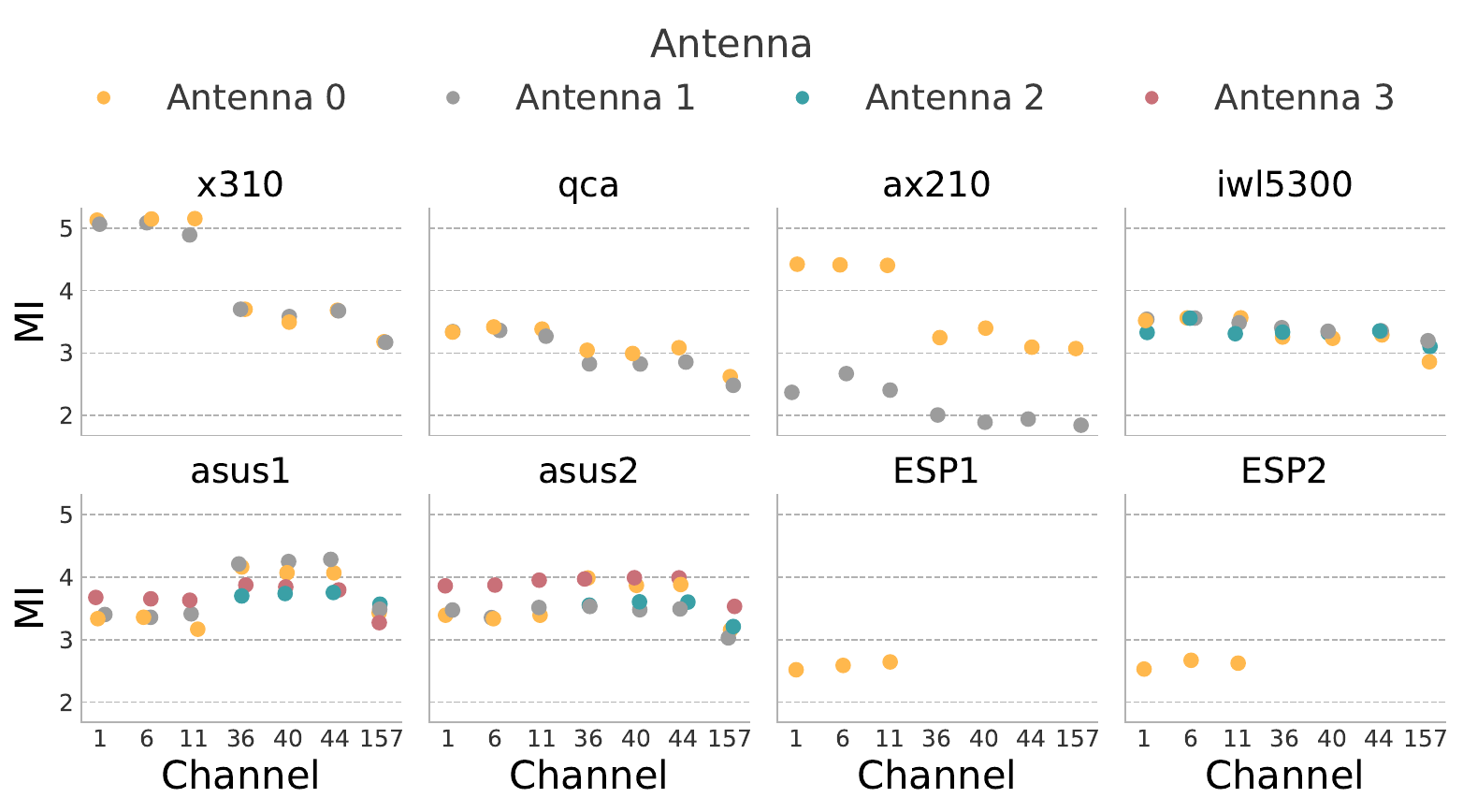}
        \Description{Amplitude sensitivity. Higher is better.}
    }

    \caption{Median (across subcarriers) mean (across time) amplitude response deviation and sensitivities for different channels ($1,6,11$ in $2.4$ GHz, $36,40,44,157$ in $5$ GHz bands) and antennas.}
    \label{fig:channel-dep-prd-and-sensitivity}
    \Description{Median (across subcarriers) mean (across time) amplitude response deviation and sensitivities for different channels and antennas.}
\end{figure}

We next investigate how faithfulness and sensitivity depend on wireless channel and receive chain. For each device, we repeat the experiment across multiple channels, capturing an identical signal simultaneously on all receive chains. For clarity, we report the subcarrier median.

The results in \Cref{fig:channel-dep-abs-prd} and \Cref{fig:channel-dep-abs-sensitivity} show minimal variation across most channels and chains. \asus{} routers exhibit a clear systematic deviation of faithfulness in channel $157$---their reported values consistently differ from expectation. Crucially, sensitivity does not drop, indicating a systematic (informationally irrelevant) offset rather than signal degradation.

In contrast, \usrp{} and \intelax{} show a marked sensitivity decrease on the $5$\,GHz channels ($36, 40, 44, 157$), \qca{} shows a slight reduction, and other devices remain stable. Antenna-wise, \asus{} devices show selective chain differences by channel. Most notably, the two \intelax{} receive chains diverge sharply, with the second underperforming; the \intelax{} labels its antennas \emph{primary} and \emph{secondary}, possibly suggesting differing capabilities.

Although channel- and antenna-dependent effects are subtle, they matter for interpreting measurements. Despite reduced high-channel sensitivity, \usrp{} remains superior or on par overall. Phase sensitivity and faithfulness show the same qualitative trends (omitted for brevity).

\section{Conclusion} \label{sec:conclusion}

In this work, we systematically investigated the impact of receiver-specific effects on \wifi{} \gls{csi} across a wide range of off-the-shelf hardware platforms. Using a combination of controlled cable-coupled experiments with precisely engineered precoded signals and \gls{ota} evaluations, we successfully disentangled receiver-induced artifacts from inherent wireless channel characteristics. Our \gls{ota} experiments specifically demonstrated how hardware-dependent differences, primarily driven by \gls{agc}, impose substantial variability that severely limits cross-device portability of sensing models.

Collectively, our insights lead to a practical, hardware-agnostic preprocessing pipeline: acquiring raw \gls{csi}, applying $\ell_{1}$ gain removal with feature-wise standardization, and calibrating stable subcarrier nonlinearities. Validated across four human activity recognition architectures and against a range of existing gain preprocessing and standardization schemes, this pipeline significantly enhances device agnosticism and supports practical, robust, and portable \wifi{} sensing deployments.

\paragraph*{Future Work} To further enhance hardware independence, future work should explore adaptive, self-supervised preprocessing capable of mitigating receiver-specific distortions in situ. Eliminating reference cables through cable-free calibration could also streamline deployment.

Even marginal improvements in recovering large-scale gain shifts would advance plug-and-play sensing. Standardizing \gls{csi} reporting, \eg normalized value ranges and explicit \gls{agc} reporting, in the IEEE 802.11bf amendment would promote robust, cross-device interoperability.

Another open question is how individual hardware aspects, for example effective \gls{csi} bits, dynamic range limits, and filter responses, each shape the observed mismatch between receivers. If these contributions can be quantified, basic hardware descriptors might be sufficient to design informed rules for transferring models across devices.

\section*{Data and Code Availability}
Our \gls{csi} measurements, derived data, and the full collection and processing pipeline used in this paper are publicly available. The datasets, containing \gls{har} and over-the-cable data, as well as resulting figures, are archived at \url{https://doi.org/10.4121/48e1331b-91df-4e1a-906b-b0a0e9d16c74}, and the collection, neural network training and processing scripts are available at \url{https://github.com/nzqo/sensession}.

\section{acknowledgment}
This research has been partially funded by the European Union’s EU Framework Programme for Research and Innovation under the HORIZON-MSCA-DN-2022 Grant Agreement No. 101119652 (MSCA-DN-6th Sense). This work has been partially supported by DFG SenShield (447586980) and partially funded by the LOEWE initiative (Hesse, Germany) within the emergenCITY center [LOEWE/1/12/519/03/05.001(0016)/72]. This work was further partially supported by the European Union - Next Generation EU under the Italian National Recovery and Resilience Plan (NRRP), Mission 4, Component 2, Investment 1.3, CUP E63C22002070006, partnership on “Telecommunications of the Future” (PE00000001 - program “RESTART”).

\bibliographystyle{IEEEtran}
\bibliography{reference}

@book{tse2005fundamentals,
author = {Tse, David and Viswanath, Pramod},
title = {Fundamentals of wireless communication},
year = {2005},
isbn = {0521845270},
publisher = {Cambridge University Press},
address = {USA}
}

@inproceedings{wisleep2014,
  author    = {Liu, Xuefeng and Cao, Jiannong and Tang, Shaojie and Wen, Jiaqi},
  title     = {Wi-Sleep: Contactless Sleep Monitoring via WiFi Signals},
  booktitle = {Proceedings of the 35th IEEE Real-Time Systems Symposium (RTSS)},
  year      = {2014},
  pages     = {346--355},
  publisher = {IEEE Computer Society},
  address   = {Piscataway, NJ, USA},
  doi       = {10.1109/RTSS.2014.30},
}

@article{wang2016csi,
  title={CSI phase fingerprinting for indoor localization with a deep learning approach},
  author={Wang, Xuyu and Gao, Lingjun and Mao, Shiwen},
  journal={IEEE Internet of Things Journal},
  volume={3},
  number={6},
  pages={1113--1123},
  year={2016},
  publisher={IEEE}
}

@inproceedings{wisign2017,
  author    = {Shang, Jiacheng and Wu, Jie},
  title     = {A Robust Sign Language Recognition System with Multiple Wi-Fi Devices},
  booktitle = {Proceedings of the Workshop on Mobility in the Evolving Internet Architecture},
  series    = {MobiArch ’17},
  location  = {Los Angeles, CA, USA},
  pages     = {19--24},
  numpages  = {6},
  year      = {2017},
  isbn      = {9781450350594},
  publisher = {Association for Computing Machinery},
  address   = {New York, NY, USA},
  url       = {https://doi.org/10.1145/3097620.3097624},
  doi       = {10.1145/3097620.3097624},
}

@inproceedings{gaussian2020,
  author    = {Rocamora, Josyl Mariela and Ho, Ivan Wang-Hei and Mak, Man-Wai},
  title     = {Gaussian Models for CSI Fingerprinting in Practical Indoor Environment Identification},
  booktitle = {Proceedings of the IEEE Global Communications Conference (GLOBECOM 2020)},
  year      = {2020},
  pages     = {1--6},
  publisher = {Institute of Electrical and Electronics Engineers (IEEE)},
  address   = {Piscataway, NJ, USA},
  doi       = {10.1109/GLOBECOM42002.2020.9322189},
}

@article{zeng2018fullbreathe,
  title={FullBreathe: Full human respiration detection exploiting complementarity of CSI phase and amplitude of WiFi signals},
  author={Zeng, Youwei and Wu, Dan and Gao, Ruiyang and Gu, Tao and Zhang, Daqing},
  journal={Proceedings of the ACM on Interactive, Mobile, Wearable and Ubiquitous Technologies},
  volume={2},
  number={3},
  pages={1--19},
  year={2018},
  publisher={ACM New York, NY, USA}
}

@article{aril,
  author={Wang, Fei and Feng, Jianwei and Zhao, Yinliang and Zhang, Xiaobin and Zhang, Shiyuan and Han, Jinsong},
  journal={IEEE Access}, 
  title={Joint Activity Recognition and Indoor Localization With WiFi Fingerprints}, 
  year={2019},
  volume={7},
  number={},
  pages={80058-80068},
  doi={10.1109/ACCESS.2019.2923743}
}

@article{reconformer,
  author={Shang, Miao and Hong, Xiaopeng},
  journal={IEEE/CAA Journal of Automatica Sinica}, 
  title={Recurrent ConFormer for WiFi Activity Recognition}, 
  year={2023},
  volume={10},
  number={6},
  pages={1491-1493},
  doi={10.1109/JAS.2023.123291}
}

@article{wiadn,
  author={Zhou, Jinggan and Liao, Xuewen and Gao, Zhenzhen and Li, Qiao and Zheng, Chunlei},
  journal={IEEE Internet of Things Journal}, 
  title={WiADN: Asymmetrical Dual-Task Attention Network for WiFi Sensing}, 
  year={2024},
  volume={11},
  number={24},
  pages={39435-39447},
  doi={10.1109/JIOT.2024.3434738}
}

@article{rganet,
    author = {Hu, Jianyuan and Ge, Fei and Cao, Xinyu and Yang, Zhimin},
    title = {RGANet: A Human Activity Recognition Model for Extracting Temporal and Spatial Features from WiFi Channel State Information},
    journal = {Sensors},
    volumne = {25},
    year = {2025},
    number = {3},
    article-number = {918},
    url = {https://www.mdpi.com/1424-8220/25/3/918},
    PubMedID = {39943556},
    issn = {1424-8220},
    doi = {10.3390/s25030918}
}

@article{gao2020crisloc,
  title={CRISLoc: Reconstructable CSI fingerprinting for indoor smartphone localization},
  author={Gao, Zhihui and Gao, Yunfan and Wang, Sulei and Li, Dan and Xu, Yuedong},
  journal={IEEE Internet of Things Journal},
  volume={8},
  number={5},
  pages={3422--3437},
  year={2020},
  publisher={IEEE}
}

@article{liu2021wiphone,
  title={WiPhone: Smartphone-based respiration monitoring using ambient reflected WiFi signals},
  author={Liu, Jinyi and Zeng, Youwei and Gu, Tao and Wang, Leye and Zhang, Daqing},
  journal={Proceedings of the ACM on Interactive, Mobile, Wearable and Ubiquitous Technologies},
  volume={5},
  number={1},
  pages={1--19},
  year={2021},
  publisher={ACM New York, NY, USA}
}

@article{cominelli2024physical,
  author={Cominelli, Marco and Shahcheraghi, Shaghayegh and Link, Jakob and Hollick, Matthias and Cerutti, Federico and Gringoli, Francesco and Asadi, Arash},
  journal={IEEE Transactions on Wireless Communications}, 
  title={Physical-Layer Privacy via Randomized Beamforming Against Adversarial Wi-Fi Sensing: Analysis, Implementation, and Evaluation}, 
  year={2024},
  volume={23},
  number={12},
  pages={19603-19617},
  doi={10.1109/TWC.2024.3485477}
}

@article{wu2024sensing,
    author={Wu, Kai and Pegoraro, Jacopo and Meneghello, Francesca and Zhang, J. Andrew and Lacruz, Jesus O. and Widmer, Joerg and Restuccia, Francesco and Rossi, Michele and Huang, Xiaojing and Zhang, Daqing and Caire, Giuseppe and Guo, Y. Jay},
    journal={IEEE Signal Processing Magazine}, 
    title={Sensing in Bistatic ISAC Systems With Clock Asynchronism: A signal processing perspective}, 
    year={2024},
    volume={41},
    number={5},
    pages={31-43},
    doi={10.1109/MSP.2024.3418725},
}

@inproceedings{hernesp32,
author = { Hernandez, Steven M. and Bulut, Eyuphan },
booktitle = { 2020 IEEE 21st International Symposium on "A World of Wireless, Mobile and Multimedia Networks" (WoWMoM) },
title = {{ Lightweight and Standalone IoT Based WiFi Sensing for Active Repositioning and Mobility }},
year = {2020},
volume = {},
ISSN = {},
pages = {277-286},
doi = {10.1109/WoWMoM49955.2020.00056},
url = {https://doi.ieeecomputersociety.org/10.1109/WoWMoM49955.2020.00056},
publisher = {IEEE Computer Society},
address = {Los Alamitos, CA, USA},
month =sep
}

@inproceedings{nexmon,
author = {Gringoli, Francesco and Schulz, Matthias and Link, Jakob and Hollick, Matthias},
title = {Free Your CSI: A Channel State Information Extraction Platform For Modern Wi-Fi Chipsets},
year = {2019},
isbn = {9781450369312},
publisher = {Association for Computing Machinery},
address = {New York, NY, USA},
url = {https://doi.org/10.1145/3349623.3355477},
doi = {10.1145/3349623.3355477},
booktitle = {Proceedings of the 13th International Workshop on Wireless Network Testbeds, Experimental Evaluation \& Characterization},
pages = {21–28},
numpages = {8},
location = {Los Cabos, Mexico},
series = {WiNTECH '19}
}

@article{halperincsitool,
	Author = {Daniel Halperin and Wenjun Hu and Anmol Sheth and David Wetherall},
	Journal = {ACM SIGCOMM CCR},
	Month = {Jan.},
	Number = {1},
	Pages = {53},
	Title = {Tool Release: Gathering 802.11n Traces with Channel State Information},
	Volume = {41},
	Year = {2011}
}

@article{jiang2021eliminating,
  title={Eliminating the barriers: Demystifying wi-fi baseband design and introducing the picoscenes wi-fi sensing platform},
  author={Jiang, Zhiping and Luan, Tom H and Ren, Xincheng and Lv, Dongtao and Hao, Han and Wang, Jing and Zhao, Kun and Xi, Wei and Xu, Yueshen and Li, Rui},
  journal={IEEE Internet of Things Journal},
  volume={9},
  number={6},
  pages={4476--4496},
  year={2021},
  publisher={IEEE}
}

@inproceedings{qian2014pads,
    author={Qian, Kun and Wu, Chenshu and Yang, Zheng and Liu, Yunhao and Zhou, Zimu},
    title={PADS: Passive detection of moving targets with dynamic speed using PHY layer information},
    booktitle={2014 20th IEEE International Conference on Parallel and Distributed Systems (ICPADS)},
    year={2014},
    isbn={9781479976157},
    issn={1521-9097},
    doi={10.1109/PADSW.2014.7097784},
    language = {eng},
    volume={},
    number={},
    address={Piscataway, NJ, USA},
    pages={1-8},
    publisher={IEEE},
}

@inproceedings{xie2015precise,
     author = {Xie, Yaxiong and Li, Zhenjiang and Li, Mo},
     title = {Precise Power Delay Profiling with Commodity WiFi},
     booktitle = {Proceedings of the 21st Annual International Conference on Mobile Computing and Networking},
     series = {MobiCom '15},
     year = {2015},
     isbn = {978-1-4503-3619-2},
     location = {Paris, France},
     pages = {53–64},
     numpages = {12},
     url = { http://doi.acm.org/10.1145/2789168.2790124 },
     doi = {10.1145/2789168.2790124},
     acmid = {2790124},
     publisher = {ACM},
     address = {New York, NY, USA},
}

@inproceedings{zhuo2016ident,
    author={Zhuo, Yiwei and Zhu, Hongzi and Xue, Hua},
    booktitle={2016 IEEE 22nd International Conference on Parallel and Distributed Systems (ICPADS)}, 
    title={Identifying a New Non-Linear CSI Phase Measurement Error with Commodity WiFi Devices}, 
    year={2016},
    volume={},
    number={},
    pages={72-79},
    doi={10.1109/ICPADS.2016.0019},
    publisher={IEEE},
    address={Piscataway, NJ, USA},
}

@article{li2017indotrack,
  title={IndoTrack: Device-free indoor human tracking with commodity Wi-Fi},
  author={Li, Xiang and Zhang, Daqing and Lv, Qin and Xiong, Jie and Li, Shengjie and Zhang, Yue and Mei, Hong},
  journal={Proceedings of the ACM on Interactive, Mobile, Wearable and Ubiquitous Technologies},
  volume={1},
  number={3},
  pages={1--22},
  year={2017},
  publisher={ACM New York, NY, USA}
}

@article{zhu2018pi,
    title={$\pi$-splicer: Perceiving accurate CSI phases with commodity WiFi devices},
    author={Zhu, Hongzi and Zhuo, Yiwei and Liu, Qinghao and Chang, Shan},
    journal={IEEE Transactions on Mobile Computing},
    volume={17},
    number={9},
    pages={2155--2165},
    year={2018},
    publisher={IEEE}
}

@article{chen2019residual,
  title={Residual carrier frequency offset estimation and compensation for commodity WiFi},
  author={Chen, Yan and Su, Xiang and Hu, Yang and Zeng, Bing},
  journal={IEEE Transactions on Mobile Computing},
  volume={19},
  number={12},
  pages={2891--2902},
  year={2019},
  publisher={IEEE}
}

@article{zhang2019calibrating,
  title={Calibrating phase offsets for commodity WiFi},
  author={Zhang, Dongheng and Hu, Yang and Chen, Yan and Zeng, Bing},
  journal={IEEE Systems Journal},
  volume={14},
  number={1},
  pages={661--664},
  year={2019},
  publisher={IEEE}
}

@inproceedings{vogt2019precise,
    author={Vogt, Hendrik and Li, Chu and Sezgin, Aydin and Zenger, Christian},
    booktitle={2019 IEEE International Workshop on Information Forensics and Security (WIFS)}, 
    title={On the Precise Phase Recovery for Physical-Layer Authentication in Dynamic Channels}, 
    year={2019},
    volume={},
    number={},
    pages={1-6},
    doi={10.1109/WIFS47025.2019.9034987},
    address={Piscataway, NJ, USA},
    publisher={IEEE},
}

@article{tadayon2019decimeter,
  title={Decimeter ranging with channel state information},
  author={Tadayon, Navid and Rahman, Muhammed Tahsin and Han, Shuo and Valaee, Shahrokh and Yu, Wei},
  journal={IEEE Transactions on Wireless Communications},
  volume={18},
  number={7},
  pages={3453--3468},
  year={2019},
  publisher={IEEE}
}

@inproceedings{zubow2021phase,
  author       = {Zubow, Anatolij and Gaw{\l}owicz, Piotr and Dressler, Falko},
  title        = {On Phase Offsets of 802.11 ac Commodity WiFi},
  booktitle    = {2021 16th Annual Conference on Wireless On-demand Network Systems and Services (WONS)},
  year         = {2021},
  pages        = {1--4},
  organization = {IEEE},
  doi          = {10.23919/WONS51326.2021.9415548},
  address      = {Klosters, Switzerland},
  publisher    = {IEEE},
}

@article{niu2021wimonitor,
  title={WiMonitor: Continuous long-term human vitality monitoring using commodity Wi-Fi devices},
  author={Niu, Xiaopeng and Li, Shengjie and Zhang, Yue and Liu, Zhaopeng and Wu, Dan and Shah, Rahul C and Tanriover, Cagri and Lu, Hong and Zhang, Daqing},
  journal={Sensors},
  volume={21},
  number={3},
  pages={751},
  year={2021},
  publisher={MDPI}
}

@inproceedings{li2021kalman,
  author    = {Li, Chu and Brauer, Jeremy and Sezgin, Aydin and Zenger, Christian},
  title     = {Kalman Filter Based MIMO CSI Phase Recovery for COTS WiFi Devices},
  booktitle = {ICASSP 2021 - 2021 IEEE International Conference on Acoustics, Speech and Signal Processing (ICASSP)},
  year      = {2021},
  pages     = {4820--4824},
  publisher = {IEEE},
  address   = {2711 Pierre Place, College Station, Texas 77845, USA},
  doi       = {10.1109/ICASSP39728.2021.9413408}
}

@inproceedings{widrone,
    author    = {Chi, Guoxuan and Yang, Zheng and Xu, Jingao and Wu, Chenshu and Zhang, Jialin and Liang, Jianzhe and Liu, Yunhao},
    title     = {Wi-drone: wi-fi-based 6-DoF tracking for indoor drone flight control},
    year      = {2022},
    isbn      = {9781450391856},
    publisher = {Association for Computing Machinery},
    address   = {New York, NY, USA},
    url       = {https://doi.org/10.1145/3498361.3538936},
    doi       = {10.1145/3498361.3538936},
    booktitle  = {Proceedings of the 20th Annual International Conference on Mobile Systems, Applications and Services},
    pages      = {56–68},
    numpages   = {13},
    location   = {Portland, Oregon},
    series     = {MobiSys '22}
}

@article{chen2023,
author = {Chen, Ching-Lan and Ko, Chun-Hsien and Wu, Sau-Hsuan and Tseng, Heng-Shih and Chang, Ronald Y.},
title = {Device-Free Target Following with Deep Spatial and Temporal Structures of CSI},
year = {2023},
issue_date = {Nov 2023},
publisher = {Kluwer Academic Publishers},
address = {USA},
volume = {95},
number = {11},
issn = {1939-8018},
url = {https://doi.org/10.1007/s11265-023-01862-y},
doi = {10.1007/s11265-023-01862-y},
journal = {J. Signal Process. Syst.},
month = jun,
pages = {1327–1340},
numpages = {14}
}

@article{yi2024enabling,
    author = {Yi, Enze and Zhang, Fusang and Xiong, Jie and Niu, Kai and Yao, Zhiyun and Zhang, Daqing},
    title = {Enabling WiFi Sensing on New-generation WiFi Cards},
    year = {2024},
    issue_date = {December 2023},
    publisher = {Association for Computing Machinery},
    address = {New York, NY, USA},
    volume = {7},
    number = {4},
    url = {https://doi.org/10.1145/3633807},
    doi = {10.1145/3633807},
    journal = {Proceedings of the ACM on Interactive, Mobile, Wearable and Ubiquitous Technologies},
    month = jan,
    articleno = {205},
    numpages = {26}
}

@article{ratnam2024optimal,
  author={Ratnam, Vishnu V. and Chen, Hao and Chang, Hao-Hsuan and Sehgal, Abhishek and Zhang, Jianzhong},
  journal={IEEE Transactions on Wireless Communications}, 
  title={Optimal Preprocessing of WiFi CSI for Sensing Applications}, 
  year={2024},
  volume={23},
  number={9},
  pages={10820-10833},
  doi={10.1109/TWC.2024.3376332},
  publisher={IEEE},
}

@book{lehmann2006theory,
  title     = {Theory of Point Estimation},
  author    = {Lehmann, Erich L. and Casella, George},
  edition   = {2},
  series    = {Springer Texts in Statistics},
  volume    = {57},
  publisher = {Springer Science \& Business Media},
  address   = {New York, NY},
  year      = {2006},
  isbn      = {978-0-387-22728-3},
  doi       = {10.1007/b98854},
}

@article{efron1987,
    issn = {01621459, 1537274X},
    url = {http://www.jstor.org/stable/2289144},
    author = {Bradley Efron},
    journal = {Journal of the American Statistical Association},
    number = {397},
    pages = {171--185},
    publisher = {[American Statistical Association, Taylor & Francis, Ltd.]},
    title = {Better Bootstrap Confidence Intervals},
    urldate = {2025-04-29},
    volume = {82},
    year = {1987}
}

@article{kim2009estimating,
    title        = {Estimating Classification Error Rate: Repeated Cross‐Validation, Repeated Hold‐Out and Bootstrap},
    author       = {Kim, Ji‐Hyun},
    journal      = {Computational Statistics \& Data Analysis},
    volume       = {53},
    number       = {11},
    issn         = {0167-9473},
    doi          = {https://doi.org/10.1016/j.csda.2009.04.009},
    pages        = {3735--3745},
    year         = {2009},
    url          = {https://www.sciencedirect.com/science/article/pii/S0167947309001601}
}

@inproceedings{nounbiasedkfold,
    author    = {Bengio, Yoshua and Grandvalet, Yves},
    booktitle = {Advances in Neural Information Processing Systems},
    editor    = {S. Thrun and L. Saul and B. Sch\"{o}lkopf},
    pages     = {},
    publisher = {MIT Press},
    address   = {Cambridge, MA, USA},
    title     = {No Unbiased Estimator of the Variance of K-Fold Cross-Validation},
    url       = {https://proceedings.neurips.cc/paper_files/paper/2003/file/e82c4b19b8151ddc25d4d93baf7b908f-Paper.pdf},
    volume    = {16},
    year      = {2003}
}

@article{kfoldrecommendation,
    title    = {Don’t lose samples to estimation},
    journal  = {Patterns},
    volume   = {3},
    number   = {12},
    pages    = {100612},
    year     = {2022},
    issn     = {2666-3899},
    doi      = {https://doi.org/10.1016/j.patter.2022.100612},
    url      = {https://www.sciencedirect.com/science/article/pii/S2666389922002379},
    author   = {Ioannis Tsamardinos}
}

@article{kraskov2004estimating,
  title = {Estimating mutual information},
  author = {Kraskov, Alexander and St\"ogbauer, Harald and Grassberger, Peter},
  journal = {Phys. Rev. E},
  volume = {69},
  issue = {6},
  pages = {066138},
  numpages = {16},
  year = {2004},
  month = {Jun},
  publisher = {American Physical Society},
  doi = {10.1103/PhysRevE.69.066138},
  url = {https://link.aps.org/doi/10.1103/PhysRevE.69.066138}
}

\appendix{}
\glsresetall[acronym]

\subsection{Effect of phase sanitization methods} \label{subsec:pads-phase-comp}

In \Cref{subsec:doppler}, we remove linear phase ramps by subtracting a \gls{ls} fit across subcarriers. An alternative, presented first in PADS~\cite{qian2014pads}, estimates the slope only from the outermost subcarriers and removes the offset as the mean. Here, we compare these two phase sanitization schemes across all receivers and tasks.

\begin{figure}[!ht]
  \centering
  \includegraphics[width=\linewidth]{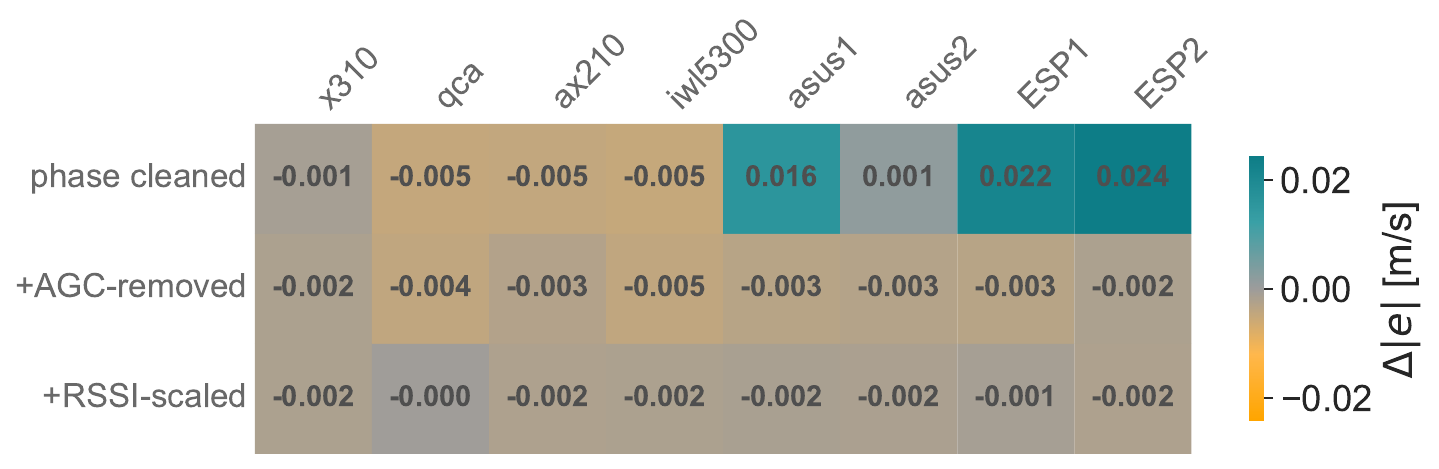}
  \caption{Change in Doppler velocity estimation error when using \gls{ls} instead of PADS. Negative values indicate that \gls{ls} yields lower error.}
  \Description{Matrix of Doppler velocity error differences between \gls{ls} and PADS phase correction, for phase cleaned, \gls{rssi}-scaled, and \gls{agc}-compensated data and for each receiver.}
  \label{fig:dopppler-phase-methods}
\end{figure}

\Cref{fig:dopppler-phase-methods} summarizes the impact of \gls{ls} versus PADS phase sanitization on the Doppler-MUSIC velocity estimates. Each matrix entry shows the change in Doppler velocity estimation error when replacing PADS by \gls{ls} for a given receiver and preprocessing variant (phase cleaned, \gls{rssi}-scaled, or \gls{agc}-compensated data). Negative values indicate that \gls{ls} yields lower error than PADS. Once large-scale gain effects have been handled (\gls{rssi} scaling or \gls{agc} compensation), the \gls{ls} method consistently outperforms PADS on every receiver, typically reducing the Doppler error by about $1$--$5\,\mathrm{mm/s}$. On the data that was only phase-cleaned, without any large-scale gain correction, the difference between the two phase sanitization methods is less systematic.

\begin{figure}[!ht]
  \centering
  \includegraphics[width=\linewidth]{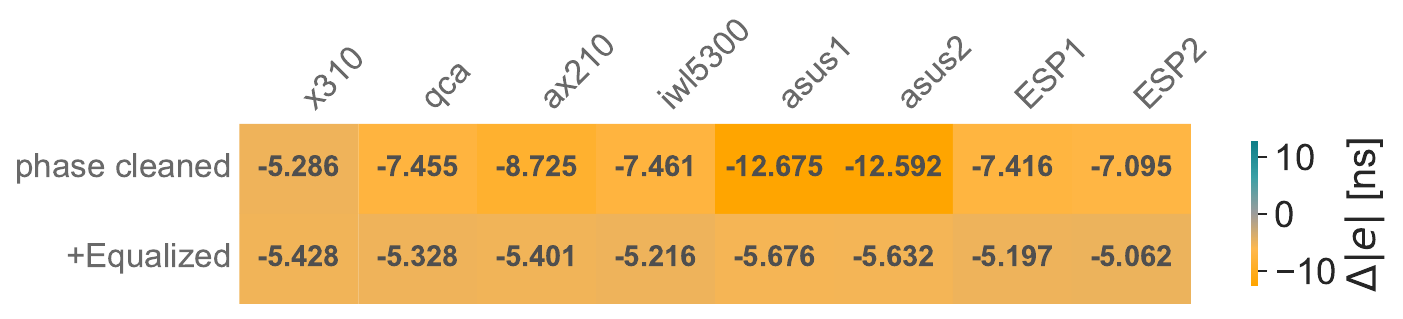}
  \caption{Change in \gls{tof} estimation error when using \gls{ls} instead of PADS. Negative values indicate that \gls{ls} yields lower error.}
  \Description{Change in \gls{tof} estimation error when using \gls{ls} phase correction instead of PADS for phase cleaned and shape-equalized data across all receivers. Negative values indicate that \gls{ls} yields lower error than PADS.}
  \label{fig:tof-phase-methods}
\end{figure}

\Cref{fig:tof-phase-methods} shows a similar comparison for the \gls{tof} estimates from \Cref{subsec:pdp-tof}. Here, we consider phase-cleaned and shape-equalized data and report the change in \gls{tof} estimation error when switching from the PADS slope estimate to \gls{ls}. With the PADS slope estimate on equalized data, all \gls{tof} errors stay on the order of $5$--$6\,\mathrm{ns}$ across receivers. Switching to the \gls{ls} fit reduces these errors by approximately $5$--$5.7\,\mathrm{ns}$, bringing the best receiver (x310) down to $0.12\,\mathrm{ns}$ and the NIC-class cards into the $0.2$--$0.7\,\mathrm{ns}$ range, with only the ESP boards still slightly above $1\,\mathrm{ns}$. Thus, \gls{ls} phase sanitization is an important enabler for the high absolute accuracies we report, while the remaining performance spread is still dominated by receiver-dependent effects.

\end{document}